\documentclass[preprint2,6pt]{emulateapj}
\usepackage{graphicx}
\usepackage{amsmath}
\usepackage{amsfonts}
\usepackage{amssymb}%
\providecommand{\U}[1]{\protect\rule{.1in}{.1in}}

\newcommand{\beq}{\begin{equation}}
\newcommand{\eeq}{\end{equation}}
\newcommand{\ba}{\begin{array}}
\newcommand{\ea}{\end{array}}

\begin{document}

\title{Hydrodynamic Properties of Gamma-Ray Burst Outflows Deduced From the Thermal Component}
\shorttitle{}
\author{Asaf Pe'er \altaffilmark{1}, Hugh Barlow \altaffilmark{1}, Shane O'Mahony \altaffilmark{1},  Raffaella Margutti \altaffilmark{2}, Felix Ryde  \altaffilmark{3}, Josefin Larsson \altaffilmark{3}, Davide Lazzati  \altaffilmark{4}, Mario Livio \altaffilmark{5}   }
\shortauthors{Pe\'er et. al.}

\altaffiltext{1}{Physics Department, University College Cork, Cork, Ireland}
\altaffiltext{2}{Harvard-Smithsonian Center for Astrophysics, 60 Garden Street, Cambridge, MA 02138, USA}
\altaffiltext{3}{The Oskar Klein Centre for Cosmoparticle Physics, SE-106 91 Stockholm, Sweden; Department of Physics, KTH Royal Institute of Technology, AlbaNova, SE-106 91 Stockholm, Sweden}
\altaffiltext{4}{Department of Physics, Oregon State University, 301 Weniger Hall, Corvallis, OR 97331, USA}
\altaffiltext{5}{28 Stablemere Ct., Baltimore, MD 21209, USA}

\begin{abstract}

We study the properties of a significant thermal emission component
that was identified in 47 GRBs observed by different
instruments. Within the framework of the ``fireball'' model, we deduce
the values of the Lorentz factor $\Gamma$, and the acceleration
radius, $r_0$, for these bursts. We find that all the values of
$\Gamma$ in our sample are in the range $10^2 \leq \Gamma \leq 10^3$,
with $\langle \Gamma \rangle = 370$.  We find a very weak dependence
of $\Gamma$ on the acceleration radius $r_0$, $\Gamma \propto
r_0^\alpha$ with $\alpha = -0.10 \pm 0.09$ at the $\sigma = 2.1$
confidence level.  The values of $r_0$ span a wide range, $10^{6.5} \leq
r_0 \leq 10^{9.5}$~cm, with a mean value of $\langle r_0 \rangle \sim
10^{8}$~cm. This is higher than the gravitational radius of a $10
M_\odot$ black hole by a factor $\approx 30$. We argue that this
result provides indirect evidence for jet propagation inside a
massive star, and suggests the existence of recollimation shocks that
take place close to this radius.

\end{abstract}
\maketitle

\section{Introduction}

One of the major developments in the study of gamma-ray bursts (GRBs)
in recent years has been the realization that a thermal component may
be a key spectral ingredient. While the shape of most GRB spectra do
not resemble a ``Planck'' function, in a non-negligible minority of
GRBs a careful spectral analysis reveals a spectral component that is
consistent with having a black-body (``Planck'') shape, accompanied by
an additional, non-thermal part.  Following pioneering work by
\citet{Ryde04, Ryde05}, such a component was clearly identified in 56
GRBs detected by BATSE \citep{RP09} and is now identified in several
Fermi GRBs as well. A few notable examples are GRB090902B
\citep{Ryde+10}, GRB100507 \citep{Ghirlanda+13}, GRB100724B
\citep{Guiriec+11}, GRB110721A \citep{Axelsson+12, Iyyani+13},
GRB120323A \citep{Guiriec+13} and GRB101219B
\citep{Larsson+15}.\footnote{A long lasting thermal component was
  identified in addition in several low luminosity GRBs, such as
  GRB060218 \citep{Campana+06} and GRB100316D \citep{Starling+11,
    Starling+12}. As these GRBs may have a different origin than
  ``classical'' GRBs \citep{Shcherbakov+13, Margutti+13}, they are
  omitted from the analysis carried out here.}

The existence of a thermal emission component may hold the key to
understanding the prompt emission spectra. First, this component
provides a physical explanation to at least part of the observed
spectra. Furthermore, thermal photons serve as seed photons for
inverse Compton (IC) scattering by energetic electrons, therefore
these photons may play an important role in explaining the non-thermal part
of the spectra as well. In fact, as it was recently shown by
\citet{AB15, Yu+15b}, while clear ``Planck'' spectra are only rarely observed,
the narrowness of the spectral width of GRBs rules out a pure
synchrotron origin in nearly 100\% of the GRB spectra observed to
date. Thus, it is possible that modified ``Planck'' spectra
contribute to the observed emission in a very large fraction of GRBs.

Thermal photons decouple from the plasma at the photosphere, which is by
definition the inner most region from which an electromagnetic signal
can reach the observer.  Therefore the properties of a Plank spectral
component directly reveal the physical conditions at the photosphere,
in those GRBs in which it can be directly identified.  This feature
is in contrast to the non-thermal spectral component, the origin of
which is still uncertain. This is due to the fact that the radiative
origin of the non-thermal component is still debatable, and its exact
emission radius is very poorly constrained, both theoretically and
observationally.

In the framework of the classical ``fireball'' model \citep{Pac86,
  RM92, RM94}, the photospheric radius, $r_{ph}$ depends only on two
free model parameters: the luminosity, $L$, and the Lorentz factor,
$\Gamma$ at the photospheric radius\footnote{This statement holds
  under the assumption that the photospheric radius $r_{ph}$ is larger
  than the saturation radius, $r_{sat}$, which is the radius at which
  all the available internal energy is converted to kinetic energy.}
\citep[e.g.,][and references therein]{Pac90, ANP91, MR00, Meszaros06}.
The observed temperature weakly depends, in addition, on a third
parameter, $r_0$, as $T^{ob} \propto r_0^{1/6}$. Here, $r_0$ is the
acceleration radius, which is the radius where the acceleration of
plasma to relativistic (kinetic) motion begins. Thus, by definition,
at this radius the bulk Lorentz factor $\Gamma(r_0) = 1$, while at
larger radii $\Gamma(r)$ increases at the expense of the internal
energy. In the classical ``fireball'' model, where the outflow expands
freely and magnetic fields are sub-dominant, the growth is linear,
$\Gamma (r) \propto r$ below the saturation radius $r_{sat}$, and
$\Gamma(r) \propto r^0$ above this radius, as all the available energy
is already in kinetic form.

The simple dependences of the temperature and photospheric radius on
the luminosity, $\Gamma$ and $r_0$, have a strong implication. As was
shown by \citet{Peer+07}, one can use the measured values of the
temperature and observed flux to directly measure $r_{ph}$ and
$\Gamma$. Using the fireball model scaling laws, one can then infer
the value of $r_0$. Therefore, using only observational quantities one
can deduce the entire (basic) dynamics of the earliest stages of the
jet evolution within the framework of the ``fireball'' model. These
dynamics, in turn, provide a strong tool in constraining models of GRB
progenitors.

In order to perform these calculations, knowledge of the distance to the GRB is
required, as the observed flux needs to be converted to luminosity.
Moreover, reliable estimates of the dynamical parameters ($\Gamma$,
$r_{ph}$ and $r_0$) rely on the assumption that the observed thermal
component is not strongly distorted \citep[e.g., by sub-photospheric
  dissipation; see][]{PMR06}. Validating this last assumption is,
nonetheless, relatively easy, as a significant thermal component $F_{th}
\lesssim F_\gamma$ that can be directly observed, necessitates that a
strong distortion does not occur \citep[see, however,][]{Ahlgren+15}.
  
Unfortunately, as of now, there are only very few GRBs which fulfill
both requirements, namely, (1) a significant thermal component is clearly
observed in their spectra, and (2) their luminosity distance is
known. These observational constraints limited, so far, the ability to
carry out a statistical study of the outflow properties derived from
photospheric emission.

Nevertheless, in recent years the number of GRBs in which a thermal
component could be clearly identified is rapidly increasing, and is
now at a few dozens. Furthermore, as we point out in the present work,
uncertainties in the redshift have, in fact, only a weak effect on the
deduced value of $r_0$. Consequentially, a good estimate of the value
of $r_0$ can be achieved even for those GRBs for which the redshift is
unknown. This enables the study of the hydrodynamical parameters in a
reasonable sample of GRBs for which a thermal component was
identified.

In this paper we analyze the existing data for those GRBs in which a
distinct thermal component was clearly identified. Using the existing
data, we deduce the values of the bulk Lorentz factor, $\Gamma$, and
the acceleration radius, $r_0$. We find an average value of $\Gamma
\approx 10^{2.5}$, which is similar to previous estimates, based on
various methods. On the other hand, we find that $\langle r_0 \rangle
\approx 10^{8}$~cm, higher than the common assumption of $\lesssim
10^7$~cm. These larger than expected $r_0$ values may be due to
propagation effects of the jet, such as recollimation shocks, mass
entrainment, or a non conical structure, and may provide indirect
evidence for the presence of a massive progenitor star.

This paper is organized as follows. In section \ref{sec:analysis} we
describe our sample selection and method of analysis. Our results are
presented in section \ref{sec:results}. We discuss our findings in
section \ref{sec:discussion}, before summarizing and concluding in
section \ref{sec:summary}.

\section{Sample and Method of Analysis}
\label{sec:analysis}

Several authors have identified a thermal emission component in
various GRBs (detected by both the BATSE instrument as well as by
Fermi satellite), and studied their properties. As was shown by
\citet{Peer+07}, clear identification of both the temperature and flux
in bursts with a known redshift $z$ (luminosity distance $d_L$), can
be used to infer the properties of the outflow within the
  framework of the ``fireball'' model \footnote{This model assumes
    that the acceleration is dominated by photon pressure, and that
    the magnetic fields are dynamically sub-dominant. For a general
    treatment of the dynamics in highly magnetized outflow, we refer
    the reader to \citet{GZ15}.} via
\beq
\Gamma = \left[ (1.06) (1+z)^2 d_L {Y F^{ob} \sigma_T \over 2 m_p c^3
    \mathcal{R}} \right]^{1/4},
\label{eq:1}
\eeq
and 
\beq
r_0 = 0.6 {d_L \over (1+z)^2} \left( {F^{ob.}_{BB} \over Y F^{ob.} }
\right)^{3/2} \mathcal{R}~~~{\rm cm}.
\label{eq:2}
\eeq
Here, $\sigma_T$ is Thomson's cross section, $\mathcal{R} \equiv
(F_{BB}^{ob}/\sigma {T^{ob}}^4)^{1/2}$, $F_{BB}^{ob}$ is the observed
black body flux, ${T^{ob}}$ is the temperature of the thermal
component, and $\sigma$ is Stefan-Boltzmann constant. The total (thermal +
non-thermal) flux is denoted by $F^{ob}$, and $Y \geq 1$ is the ratio
between the total energy released in the explosion producing the GRB
and the energy observed in $\gamma$-rays (both thermal and
non-thermal). Measurements of the exact value of $Y$ are difficult to
conduct, though estimates can be done using afterglow observations
\citep{Cenko+11, Peer+12, Wygoda+15}.

A key difficulty in identifying a thermal component and conducting the
calculations is the fact that the signal varies with time. Thus, a
time dependent analysis is required. As was shown by \citet{Ryde04,
  Ryde05} and \citet{RP09}, for bursts which show relatively smooth,
long pulses, and in which a thermal component could be identified,
both the temperature and the flux show a very typical behavior: a
broken power law in time. Typically, before $t_{brk}^{ob} \sim$~a
few~s, the temperature is roughly constant, while the thermal flux
increases roughly as $F^{ob}_{BB} \propto t^{1/3}$. At later times,
both the temperature and the flux decay, $T^{ob} \propto t^{-2/3}$ and
$F^{ob}_{BB} \propto t^{-2}$. Although the non-thermal and thermal
fluxes are often correlated, the break time does not always coincide
with the peak of the (non-thermal) flux. In many GRBs it is associated
with the beginning of the rapid decay of the pulse
\citep{RP09}. \footnote{We should emphasis the fact that a systematic
  study of the temporal behavior of GRB pulses was carried only for
  BATSE bursts. No such systematic study was carried so far for Fermi
  bursts.}

The exact interpretation of the early and late temporal behavior
is still not fully clear. An interesting finding is that during the
rise phase of the pulses (the first few seconds), the temperature is
roughly constant. This implies via Equation \ref{eq:1} that the
Lorentz factor is roughly constant (assuming that the ratio
$F_{BB}/F_{tot}$ does not vary much), though the total flux changes
substantially. This no longer holds during the decay phase. A leading
idea is that the late time behavior may be associated with a
geometrical effect of (relativistic) ``limb darkening'' \citep{Peer08,
  PR11, LPR13}, though a non-spherical jet structure may be required
\citep{DZ14}. If this is the correct interpretation, it implies that
reliable estimation of the hydrodynamic parameters can be done only if
the data are taken at the break time, or earlier. Fortunately, in
recent years enough data are available to carry out these time-dependent
calculations for a substantial number of GRBs.

In this work we collected the available data, and divided them into
three categories. In category (I) we include seven GRBs that fulfill
the entire set of conditions: (1) their redshifts are known; (2) a
thermal component was reported in the literature; and (3) a time
dependent analysis could be performed, and therefore the values of
$\Gamma$ and $r_0$ were inferred. These GRBs were analyzed in recent
years by various authors. Here, we collected the published data, and
validated it. It case of multiple-pulsed GRB, we normally picked the
first pulse.  Two of these GRBs, GRB970828 and GRB990510, were
detected by BATSE \citep{Peer+07}, and the other five were detected by
Fermi. We summarize in Table \ref{tab:T1} the derived values of the
parameters, as well as the references from which they were taken.

In category (II) there are four GRBs detected by Fermi, that both: (1)
show clear evidence of a significant thermal component, and (2) have
a well defined temporal evolution, that enables a clear identification
of the break time.\footnote{We omitted from our sample GRBs for which
  this temporal behavior could not be verified, such as, e.g.,
  GRB100507 \citep{Ghirlanda+13}.}  However, as opposed to GRBs in our
category (I) sample, the redshifts of these GRBs are unknown. The
dynamical properties of these GRBs can therefore be deduced only up to
the uncertainty in the redshift. The derived parameters of these GRBs,
as given by the various authors are also shown in Table \ref{tab:T1},
with the assumed values taken for the redshifts by the different
authors. In our analysis, we did not modify the assumed redshift, as
it was taken as the mean redshift of GRBs detected by the relevant
instrument (note that GRB120323 is a short GRB). As we show below,
this uncertainty does not affect our conclusions.

Although the thermal component observed in the spectra of GRBs in our
category (II) sample is statistically significant, it is relatively
weak: in 3/4 GRBs, the flux in the thermal component is less than 10\%
of the total flux observed in $\gamma$-rays. This poses a challenge to
the analysis method: as was shown by \citet{ZP09, Hascoet+13, GZ15},
the photospheric component is suppressed if the flow is highly
magnetized. Thus, weak thermal fluxes may be an indication for highly
magnetized outflow, in which case the dynamical calculations presented
in Equations \ref{eq:1} and \ref{eq:2} are modified \citep[see][for a
  full treatment in this case]{GZ15}. Nonetheless, the magnetization
parameter is unknown, and a weak thermal component does not
necessitate a high magnetization; it could result, e.g., from $r_{ph}
\gg r_{sat}$. We therefore decided to include these bursts in our
sample. As will be shown below, the values obtained for both $r_0$ and
$\Gamma$ for these bursts are similar to those obtained for the GRBs
in the rest of our sample, which may indicate similar dynamics.

\tabletypesize{\tiny}

\begin{deluxetable*}{clcccccccl}
 \tablecolumns{10}
 \tablewidth{0pc}
 \tablecaption{List of GRBs in our category (I): known redshift, and (II): Fermi GRBs without known redshift. The derived values of $\Gamma$ and $r_0$ in this table are done under the assumption $Y=1$.  See the text for details.}
\tablehead{ \colhead{Category} & \colhead{Burst } & \colhead{$z$} &  \colhead{$T$} & \colhead{$F_{\rm BB}$}
& \colhead{$F_{\rm BB}/F_{\rm tot}$} & \colhead{$\Gamma$} & \colhead{$r_0$} & \colhead{Reference } & \colhead{Comments } \\
\colhead{} &\colhead{} & \colhead{ } & \colhead{[keV] } &\colhead{[erg cm$^{-2}$ s$^{-1}$] } &
\colhead{ } & \colhead{} & \colhead{[cm]} & \colhead{} & \colhead{} }

\startdata
I & 970828  & 0.9578 & $78.5 \pm 4$ &                      & 0.64 & $ 305 \pm 28$  & $2.9 \pm 1.8 \times 10^8$ & $^1$ & a, b \\
I & 990510  & 1.619  & $46.5 \pm 2$ & $7.0 \times 10^{-7}$ & 0.25 & $ 384 \pm 71$  & $1.7 \pm 1.7 \times 10^8$ & $^1$ & b \\
I & 080810  & 3.355  & $62$         & $1.6\times 10^{-7}$  & 0.28 & $570 \pm (170)$ & $2.3 \pm (1.2) \times 10^8$ & $^2$ & c, d \\    
I & 090902B & 1.822  & $168$        & $1.96 \times 10^{-5}$ & 0.26 & $995 \pm 75$  & $5.2 \pm 2.3 \times 10^8$ & $^3$ & \\
I & 090926B  & 1.24   & $17.2 \pm 1$ & $3 \times 10^{-7}$    & 0.92 & $110 \pm 10$   & $4.3 \pm  0.9 \times 10^9$  & $^4$ & e\\
I & 101219B & 0.55   & $19.1 \pm 0.7$ & $8.45 \pm 0.03 \times 10^{-8}$ & $\lesssim 1$ & $138 \pm 8$ & $2.7 \pm 1.6 \times 10^7$ & $^5$ & \\ 
I & 110731A  &  2.83   & $85 \pm 5 $  &                      & 0.47 & $765 \pm 200$  & $3.46 \pm  1.1 \times 10^8$ & $^6$ & d \\
II & 100724B & (1)  & $38 \pm 4$     & $2.6 \times 10^{-7}$ & 0.04 & $325 \pm 100$ & $ 1.2 \pm 0.6 \times 10^7$  & $^7$ & \\ 
II & 110721A &  (2)  & $30$           &                     & 0.09 & $450 \pm 200$ & $ 1 \pm 0.4 \times 10^7$  & $^8$ & f \\                                  
II & 110920  &  (2)  & $61.3 \pm 0.7$ &                     & 0.3  & $442 \pm (133)$ & $ 2 \pm 1 \times 10^8$   & $^9$ & \\    
II & 120323 &  (0.5) & $11.5 \pm 1.5$ &                    & 0.05 &  $145 \pm (20)$ & $  2.6 \pm (0.9) \times 10^9$ & $^{10}$ & g \\    
  \tablenotetext{~}{$^1$ \citet{Peer+07}; $^2$ \citet{Page+09}; $^3$
  \citet{Peer+12}; $^4$ \citet{Serino+11}; $^5$ \citet{Larsson+15};
  $^6$ \citet{Basha13}; $^7$ \citet{Guiriec+11}; $^8$
  \citet{Iyyani+13}; $^9$ \citet{McGlynn+12, Iyyani+15}; $^{10}$
  \citet{Guiriec+13}.  }
\tablenotetext{a}{Data is based on published references. Some references omit data on $F_{BB}$, $dF_{BB}$ or $dT$.}
\tablenotetext{b}{GRBs 970828 and GRB990510 were detected by BATSE instruments. All other GRBs in this list were detected by Fermi-GBM.}
\tablenotetext{c}{Errors in $dr_0$, $d\Gamma$ are estimated based on the data provided in the reference.}
\tablenotetext{d}{In addition to Fermi-GBM, this GRB was detected by the Swift-BAT.}
\tablenotetext{e}{In addition to Fermi-GBM, this GRB was detected by MAXI.}
\tablenotetext{f}{Errors represent uncertainty in redshift as well.}
\tablenotetext{g}{Short GRB; Errors estimated from data provided in table 4 of \citet{Guiriec+13}.}
\enddata
\label{tab:T1}
\end{deluxetable*}

In category (III) we used 36 GRBs detected by BATSE, taken from the
sample of \citet{RP09}. This is the largest single sample of GRBs in
which a temporal analysis was carried out and a thermal component was
clearly identified. Out of 56 GRBs in the \citet{RP09} sample, we
selected those GRBs in which a clear break time in the temporal
behavior of both the temperature and flux was identified. Furthermore,
the break times of the temperature and flux evolution were consistent
with each other. Thus, although a break time was identified in all
GRBs in our sample, the sub-sample of GRBs in this category is
homogeneous as all GRBs are detected by the same instrument and
identical selection criteria was used.

The sample of our category (III) GRBs is given in Table
\ref{tab:T2}. The values of the temperature, thermal flux and ratio of
thermal to total flux are given at the break time; the derived values
of $\Gamma$, $r_0$ and the photospheric radius, $r_{ph}$ are under the
assumption of $z=1$.

\tabletypesize{\tiny}

\begin{deluxetable*}{lccccccl}
 \tablecolumns{10}
 \tablewidth{0pc}
 \tablecaption{Sample of our category III GRBs: from \citet{RP09}. See the text for details.}
\tablehead{ \colhead{Burst }  & \colhead{Trigger} &  \colhead{$T$} & \colhead{$F_{\rm BB}$}
& \colhead{$F_{\rm BB}/F_{\rm tot}$} & \colhead{$\Gamma$} & \colhead{$r_0$} & \colhead{$r_{ph}$} \\
\colhead{} &\colhead{} &  \colhead{[keV] } &\colhead{[erg cm$^{-2}$ s$^{-1}$] } &
\colhead{ } & \colhead{} & \colhead{[cm]} & \colhead{[cm] } 
}

\startdata 
910807 & 647 & $57.3 \pm 2.2$ & $ 1.7 \pm 0.3\times 10^{-6} $ & 0.64 & $257 \pm 10$ &$ 6.23 \pm 1.54\times 10^{8} $  & $ 4.88 \pm 0.60\times 10^{11} $\\
910814 & 678 & $173.1 \pm 12.6$ & $ 5.7 \pm 2.0\times 10^{-6} $ & 0.51 & $549 \pm 39$ &$ 8.96 \pm 4.68\times 10^{7} $  & $ 2.09 \pm 0.44\times 10^{11} $\\
911016 & 907 & $62.5 \pm 2.5$ & $ 9.5 \pm 2.0\times 10^{-7} $ & 0.67 & $247 \pm 10$ &$ 4.22 \pm 1.03\times 10^{8} $  & $ 2.96 \pm 0.38\times 10^{11} $\\
911031 & 973 & $59 \pm 6.9$ & $ 6.9 \pm 4.7\times 10^{-7} $ & 0.18 & $334 \pm 55$ &$ 7.00 \pm 8.33\times 10^{7} $  & $ 3.53 \pm 1.23\times 10^{11} $\\
920525 & 1625 & $148.5 \pm 10.4$ & $ 1.2 \pm 0.4\times 10^{-5} $ & 0.43 & $583 \pm 42$ &$ 1.38 \pm 0.78\times 10^{8} $  & $ 4.41 \pm 0.89\times 10^{11} $\\
920718 & 1709 & $44 \pm 2.3$ & $ 1.7 \pm 0.5\times 10^{-6} $ & 0.28 & $278 \pm 18$ &$ 3.02 \pm 1.66\times 10^{8} $  & $ 9.06 \pm 1.41\times 10^{11} $\\
921003 & 1974 & $15.9 \pm 1.3$ & $ 7.7 \pm 3.3\times 10^{-7} $ & 0.59 & $125 \pm 12$ &$ 5.08 \pm 3.14\times 10^{9} $  & $ 2.12 \pm 0.58\times 10^{12} $\\
921123 & 2067 & $56.1 \pm 1.8$ & $ 2.8 \pm 0.5\times 10^{-6} $ & 0.40 & $306 \pm 11$ &$ 4.13 \pm 1.19\times 10^{8} $  & $ 7.87 \pm 0.76\times 10^{11} $\\
921207 & 2083 & $94.8 \pm 3.2$ & $ 2.0 \pm 0.1\times 10^{-5} $ & 0.61 & $458 \pm 11$ &$ 7.37 \pm 1.48\times 10^{8} $  & $ 1.11 \pm 0.09\times 10^{12} $\\
930112 & 2127 & $111.7 \pm 7.5$ & $ 2.9 \pm 1.0\times 10^{-6} $ & 0.42 & $426 \pm 30$ &$ 1.13 \pm 0.63\times 10^{8} $  & $ 2.78 \pm 0.54\times 10^{11} $\\
930214 & 2193 & $100.7 \pm 7.3$ & $ 6.2 \pm 2.4\times 10^{-7} $ & 0.81 & $283 \pm 23$ &$ 1.73 \pm 0.62\times 10^{8} $  & $ 1.05 \pm 0.26\times 10^{11} $\\
930612 & 2387 & $40.2 \pm 4.3$ & $ 4.7 \pm 2.7\times 10^{-7} $ & 0.42 & $208 \pm 26$ &$ 4.15 \pm 3.93\times 10^{8} $  & $ 4.22 \pm 1.47\times 10^{11} $\\
940410 & 2919 & $61.4 \pm 14.1$ & $ 2.3 \pm 2.7\times 10^{-7} $ & 0.11 & $320 \pm 102$ &$ 1.46 \pm 4.03\times 10^{7} $  & $ 1.94 \pm 1.26\times 10^{11} $\\
940708 & 3067 & $73.2 \pm 9.8$ & $ 1.3 \pm 0.9\times 10^{-6} $ & 0.19 & $377 \pm 66$ &$ 5.51 \pm 8.25\times 10^{7} $  & $ 3.80 \pm 1.47\times 10^{11} $\\
941023 & 3256 & $46.5 \pm 7.5$ & $ 3.3 \pm 1.4\times 10^{-7} $ & 0.61 & $196 \pm 29$ &$ 4.31 \pm 3.29\times 10^{8} $  & $ 2.55 \pm 0.81\times 10^{11} $\\
941026 & 3257 & $55.4 \pm 6.4$ & $ 6.1 \pm 2.2\times 10^{-7} $ & 0.54 & $233 \pm 24$ &$ 3.29 \pm 2.04\times 10^{8} $  & $ 2.89 \pm 0.78\times 10^{11} $\\
941121 & 3290 & $41.1 \pm 2$ & $ 1.3 \pm 0.3\times 10^{-6} $ & 0.34 & $247 \pm 14$ &$ 4.05 \pm 1.94\times 10^{8} $  & $ 8.00 \pm 1.18\times 10^{11} $\\
950403 & 3492 & $108.5 \pm 4.1$ & $ 3.3 \pm 0.6\times 10^{-5} $ & 0.35 & $598 \pm 25$ &$ 3.02 \pm 1.03\times 10^{8} $  & $ 1.39 \pm 0.15\times 10^{12} $\\
950624 & 3648 & $28.9 \pm 2.6$ & $ 1.3 \pm 0.7\times 10^{-7} $ & 0.38 & $150 \pm 16$ &$ 3.06 \pm 2.76\times 10^{8} $  & $ 3.06 \pm 0.89\times 10^{11} $\\
950701 & 3658 & $67.6 \pm 3$ & $ 3.0 \pm 0.7\times 10^{-6} $ & 0.42 & $334 \pm 16$ &$ 3.13 \pm 1.22\times 10^{8} $  & $ 6.09 \pm 0.82\times 10^{11} $\\
951016 & 3870 & $28.2 \pm 3.8$ & $ 5.9 \pm 4.7\times 10^{-7} $ & 0.10 & $252 \pm 53$ &$ 0.94 \pm 1.70 \times 10^{8} $  & $ 1.17 \pm 0.48\times 10^{12} $\\
951102 & 3891 & $66.9 \pm 4.9$ & $ 2.8 \pm 1.0\times 10^{-6} $ & 0.23 & $380 \pm 35$ &$ 1.28 \pm 1.00\times 10^{8} $  & $ 6.76 \pm 1.42\times 10^{11} $\\
951213 & 3954 & $52.9 \pm 4.6$ & $ 9.4 \pm 4.3\times 10^{-7} $ & 0.22 & $300 \pm 34$ &$ 1.09 \pm 1.05\times 10^{8} $  & $ 4.97 \pm 1.27\times 10^{11} $\\
951228 & 4157 & $24.4 \pm 0.6$ & $ 4.5 \pm 0.7\times 10^{-7} $ & 0.98 & $128 \pm 4$ &$ 3.40 \pm 0.31\times 10^{9} $  & $ 6.96 \pm 0.68\times 10^{11} $\\
960124 & 4556 & $60.6 \pm 5.5$ & $ 2.3 \pm 0.9\times 10^{-6} $ & 0.22 & $364 \pm 39$ &$ 1.40 \pm 1.19\times 10^{8} $  & $ 7.29 \pm 1.71\times 10^{11} $\\
960530 & 5478 & $44.1 \pm 3.6$ & $ 4.7 \pm 0.4\times 10^{-7} $ & 0.38 & $220 \pm 9$ &$ 2.55 \pm 0.43\times 10^{8} $  & $ 3.78 \pm 0.37\times 10^{11} $\\
960605 & 5486 & $64.9 \pm 3.2$ & $ 1.4 \pm 0.4\times 10^{-6} $ & 0.54 & $279 \pm 15$ &$ 3.48 \pm 1.33\times 10^{8} $  & $ 3.80 \pm 0.60\times 10^{11} $\\
960804 & 5563 & $47.9 \pm 3$ & $ 2.7 \pm 0.9\times 10^{-6} $ & 0.26 & $311 \pm 25$ &$ 2.93 \pm 1.98\times 10^{8} $  & $ 1.06 \pm 0.20\times 10^{12} $\\
960912 & 5601 & $68.1 \pm 6.6$ & $ 7.6 \pm 3.8\times 10^{-7} $ & 0.33 & $300 \pm 34$ &$ 1.08 \pm 1.01\times 10^{8} $  & $ 2.70 \pm 0.77\times 10^{11} $\\
960924 & 5614 & $155.4 \pm 8$ & $ 7.3 \pm 1.0\times 10^{-4} $ & 0.94 & $820 \pm 24$ &$ 3.17 \pm 0.48\times 10^{9} $  & $ 4.45 \pm 0.65\times 10^{12} $\\
961102 & 5654 & $62.2 \pm 3.3$ & $ 8.1 \pm 2.3\times 10^{-7} $ & 0.35 & $283 \pm 17$ &$ 1.51 \pm 0.76\times 10^{8} $  & $ 3.17 \pm 0.50\times 10^{11} $\\
970223 & 6100 & $98.3 \pm 7$ & $ 3.7 \pm 1.3\times 10^{-6} $ & 0.32 & $441 \pm 35$ &$ 1.13 \pm 0.75\times 10^{8} $  & $ 4.23 \pm 0.86\times 10^{11} $\\ 
970925 & 6397 & $44.6 \pm 2.3$ & $ 7.6 \pm 2.1\times 10^{-7} $ & 0.33 & $243 \pm 15$ &$ 2.51 \pm 1.30\times 10^{8} $  & $ 5.10 \pm 0.80\times 10^{11} $\\
980306 & 6630 & $80.6 \pm 4.6$ & $ 5.5 \pm 1.1\times 10^{-6} $ & 0.84 & $334 \pm 13$ &$ 7.95 \pm 2.09\times 10^{8} $  & $ 5.83 \pm 0.82\times 10^{11} $\\
990102 & 7293 & $54.4 \pm 2.2$ & $ 7.0 \pm 1.6\times 10^{-7} $ & 0.58 & $229 \pm 10$ &$ 3.85 \pm 1.18\times 10^{8} $  & $ 3.10 \pm 0.42\times 10^{11} $\\
990102 & 7295 & $100.7 \pm 16.1$ & $ 1.4 \pm 0.6\times 10^{-6} $ & 0.68 & $330 \pm 45$ &$ 2.20 \pm 1.46\times 10^{8} $  & $ 1.90 \pm 0.64\times 10^{11} $\\
\enddata
\label{tab:T2}
\end{deluxetable*}

While the redshifts of all the GRBs in our categories (II) and (III)
are unknown, we point out that the additional uncertainty in the
estimate of $r_0$ due to the lack of a precise redshift is not very large. This is
due to the fact that $r_0 \propto d_L/(1+z)^2$, and, for the range of
redshifts typical for pre-Swift GRBs, $0.5 \lesssim z \lesssim 2.5$,
one finds $0.76 \leq d_L/{d_L}_{(z=1)} \times (2/(1+z))^2 \leq 1.06$.
A similar calculation shows that the added uncertainty in the estimate
of $\Gamma$ is $0.7 \leq \Gamma(z)/\Gamma_{(z=1)} \leq 1.75$, for $0.5
\leq z \leq 2.5$.

We estimated the additional uncertainty of GRBs in our category (III)
due to the unknown redshifts as follows.\footnote{We omitted the full
  calculation for the four GRBs in our category (II), as this sample
  is not homogeneous, and the additional error would not affect our
  final conclusion. For completeness, we do show in Figure \ref{fig:1}
  the values of $r_0$ and $\Gamma$ of these GRBs obtained for $z=1$.}
We used the pre-Swift distribution of GRB redshifts given by
\citet{Jakobsson+06}. We performed a Monte-Carlo simulation,
simulating $10^6$ GRBs drawn from this distribution, and calculated
the ratios $d_L/(1+z)^2$ and $(d_L \times(1+z)^2)^{1/4}$ of each GRB
in our simulation, normalized to these values for $z=1$. In our
calculation, we assumed a standard cosmology (flat universe with
$\Omega_m = 0.286$ and $H_0 = 69.6~\rm{km/s/Mpc}$).  Based on the
simulated results, we conclude that the average values of $r_0$ and
$\Gamma$ are $\langle r_0 \rangle / {r_0}_{z=1} = 0.863$, and $
\langle \Gamma \rangle / {\Gamma}_{z=1} = 1.117$. The standard
deviations due to the uncertain redshifts are $\sigma(r_0) = 0.2533$
and $\sigma(\Gamma) = 0.531$. These values are not surprising given
the above analysis and the fact that the mean redshift in the sample
of \citet{Jakobsson+06} is $\langle z \rangle = 1.345$.

The values of $\Gamma$ and $r_0$ presented in tables \ref{tab:T1},
\ref{tab:T2} are derived under the assumption of $Y=1$. This is in
order to be consistent with the results presented in the various
references from which the data is adopted.  However, in presenting the
results in the figures below, we adopt a somewhat higher value of
$Y=2$.  As explained above, the exact value of $Y$ is very difficult
to measure. Measurements based on afterglow observations reveal mixed
results. Some works found very high efficiency in $\gamma$-ray
production, implying $Y \lesssim$~a few \citep{Cenko+11, Peer+12}. On
the other hand, several works found inefficient radiation, implying
larger values of $Y$ \citep{SBK14, Wang+15}.  A large value of $Y$
would increase the measured value of the Lorentz factor by tens of \%,
as $\Gamma \propto Y^{1/4}$, while decreasing $r_0$ by a factor of a
few, as $r_0 \propto Y^{-3/2}$. As the highest values of $\Gamma$ we
obtain are $\gtrsim 10^3$, while the lowest value of $r_0$ is $\sim
10^{6.5}$~cm (see below), we deduce that the value of $Y$ cannot be
much greater than unity. In the analysis below, we therefore take as a
fiducial value $Y=2$.

\section{Results}
\label{sec:results}

\begin{figure}
\plotone{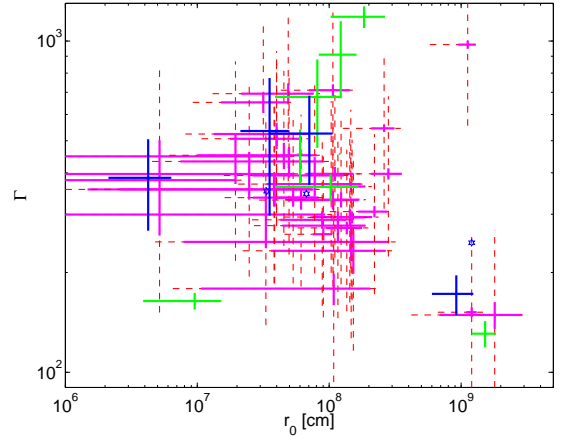}
\caption{Lorentz factor, $\Gamma$, vs acceleration radius, $r_0$, of
  GRBs in our sample. Green points are GRBs in our category (I); Blue
  points are GRBs in our category (II), while magenta points are GRBs
  from our category (III) (see text for details). Solid error bars
  represent statistical errors, while dashed error bars represent
  additional uncertainty due to unknown redshifts of GRBs in our
  category (III). The stars show the location of the parameters of
  GRBs in our category (II) with assumed redshift $z=1$. A linear fit
  reveals $\Gamma \propto r_0^{\alpha}$ with $\alpha = -0.10 \pm 0.09$
  and a very weak correlation. The values of $\Gamma$ and $r_0$
    presented are derived under the assumption $Y=2$. For a different
    value of $Y$, the results can be scaled according to $\Gamma
    \propto Y^{1/4}$ and $r_0 \propto Y^{-3/2}$.}
\label{fig:1}
\end{figure}

\begin{figure}
\plottwo{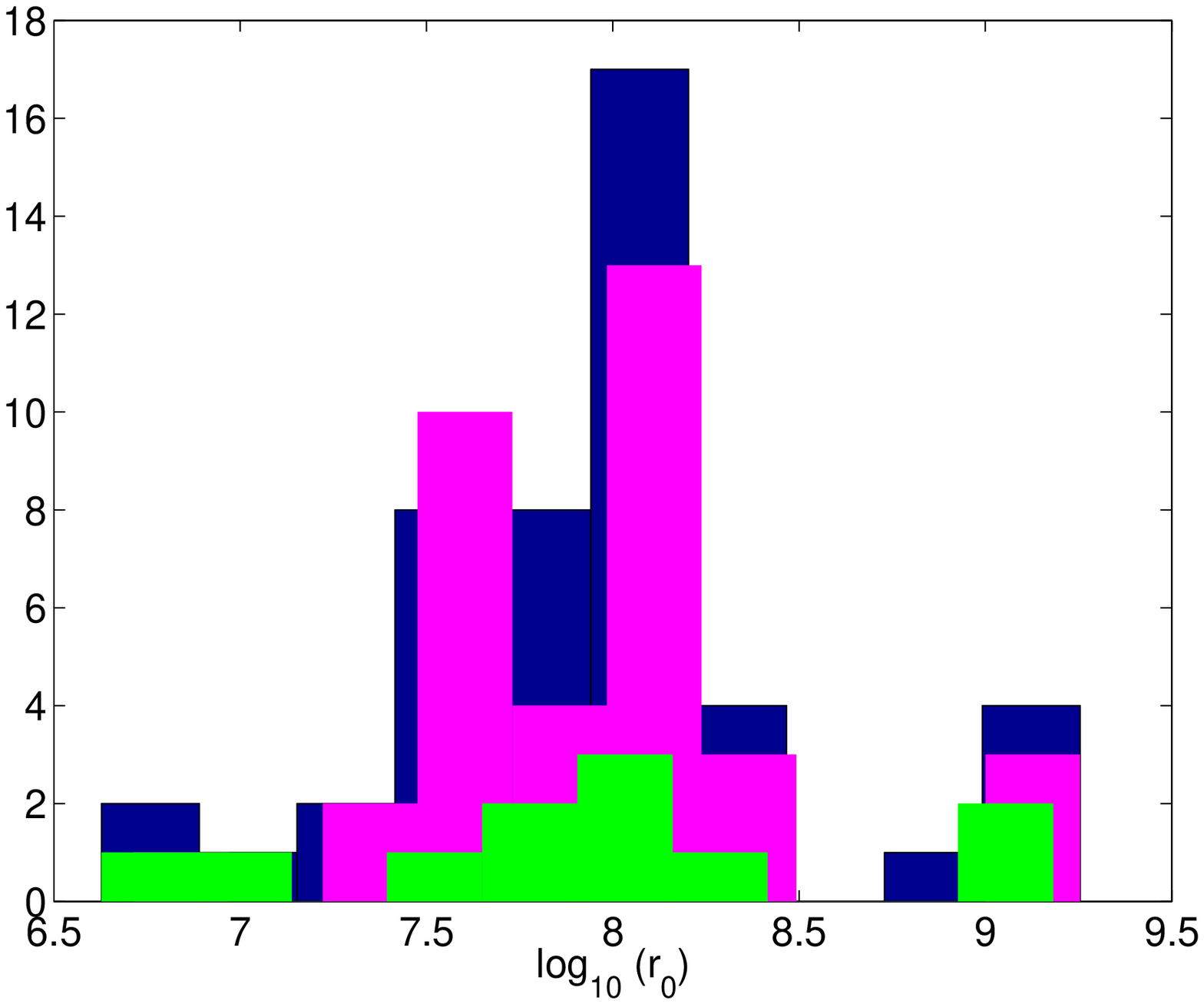}{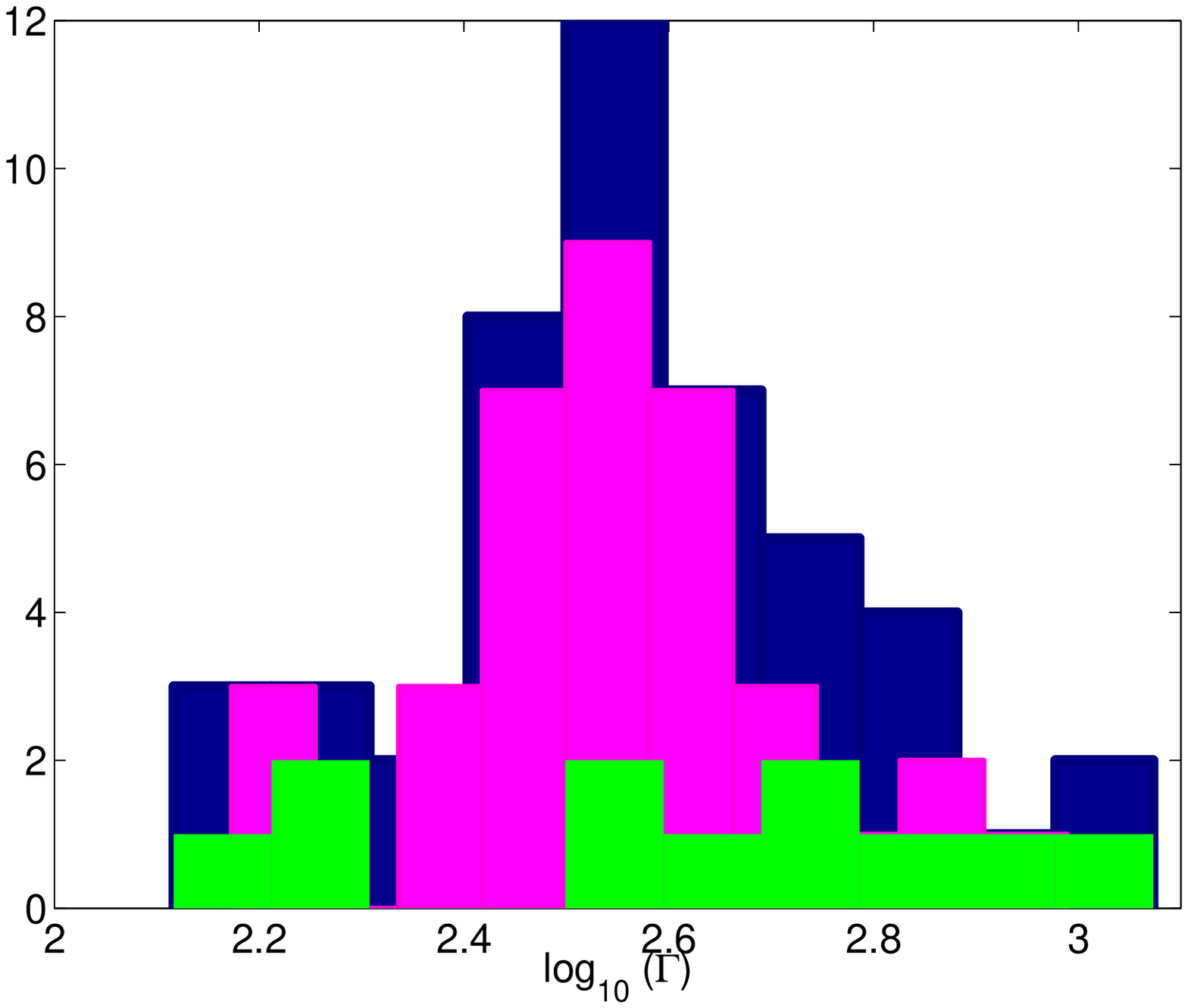}
\caption{Histograms of the mean values of $\log_{10} (r_0)$ (left) and
  $\log_{10} (\Gamma)$ (right). Blue are for the entire sample, while
  magenta are for GRBs in our category (III)
  \citep[the][]{RP09} sample only, and green are for GRBs in our categories (I) and (II). }
\label{fig:2}
\end{figure}

The inferred values of the Lorentz factor $\Gamma$ and the
acceleration radius $r_0$ from our sample are presented in Figure
\ref{fig:1}. The green points represent GRBs with known redshift (our
category I GRBs), GRBs in our category (II) are presented by the blue
points, and category (III) GRBs are shown by the magenta points. The
error bars represent statistical errors; additional errors due to the
uncertain redshifts are shown by the dashed (red) dots. The blue stars
represent the values of the parameters of GRBs in our category (II),
with assumed redshift $z=1$. As GRBs in this category form a small
fraction of our sample, the uncertainty in the redshift of these GRBs
does not affect any of our conclusions. In the plot, we assume a
fixed value of $Y=2$.  We further show in Figure \ref{fig:2}
histograms of the values of $\Gamma$ and $r_0$.

The most interesting result from these plots is the range of parameter
values. We find that the range of values of the Lorentz factor
stretches between 130 and 1200 (with 1-$\sigma$ errors inclusive),
with an average value of $\langle \Gamma \rangle = 370$, or $\langle
\log_{10} (\Gamma) \rangle = 2.57$ and standard deviation of
$\sigma(\log_{10} (\Gamma)) = 0.21$. These results are similar to the
values of $\Gamma$ inferred for bright GRBs by other methods which are
in wide use, such as opacity arguments \citep{KP91, WL95, LS01}, the
deceleration time of the emission from the forward shock \citep{SP99};
see also \citet{Racusin+11}, or the onset of the afterglow
\citep{Liang+10}. Furthermore, the results of Figure \ref{fig:2}
indicate that the distribution of $\Gamma$ is fairly well fitted by a
Gaussian, suggesting that selection biases are not playing a major
role (see further discussion below).

The initial acceleration radius, $r_0$, ranges between $4 \times 10^6
\leq r_0 \leq 2 \times 10^9$~cm. The average value is $\langle r_0
\rangle = 9.1 \times 10^7$~cm, or $\langle \log_{10} (r_0) \rangle =
7.96$ and standard deviation of $\sigma(\log_{10} (r_0)) =
0.56$. Similarly to the distribution of $\Gamma$, the distribution of
$\log_{10} (r_0)$ is also close to a Gaussian (see Figure
\ref{fig:2}).\footnote{We note that as about 75\% of the GRBs in our
  sample are in category (III), they have the dominant effect on the
  distributions.} The average value found corresponds to minimum
variability time of $\delta t = r_0/c \sim 3 \times 10^{-3}$~s, and is
about 30 times the gravitational radius of 10~$M_\odot$ black
hole. While this value is fully consistent with our data, as all the
GRBs in our sample are long, and their variability time is much longer
than 10 ms, this value is different from the commonly assumed value of
the acceleration radius, $\approx few$ gravitational radii
\citep[e.g.,][and references therein]{Meszaros06, KZ14}.

A priori, we do not expect any correlation between the values of
$\Gamma$ and of $r_0$, as these are two independent parameters of the
fireball model. Nonetheless, it is interesting to search for such a
correlation, since if it exists it could put interesting constraints on
the nature of the progenitor. This is due to the fact that if indeed
$r_0$ is related to the gravitational radius, than $r_0 \propto M$,
where $M$ is the mass of the central object, while $\Gamma \propto L/
{\dot M} c^2$, where $\dot M$ is the mass ejection rate.  

Linear fitting (with points weighted by the errors in both directions)
reveals best fitted values of $\log_{10} (\Gamma) = a_1 + a_2
\log_{10} (r_0)$, with $a_1 = 3.31 \pm 0.66$ and $a_2 = -0.10 \pm
0.08$. The correlation is thus very weak, with intrinsic scatter $0.44
\pm 0.05$ (Spearman rank -0.31 with significance of 0.035, equivalent
to $2.1 \sigma$). While this analysis is done on a non-homogeneous
sample, a similar analysis carried out on GRBs in our category (III)
only, which do form a homogeneous sample, is not significantly
different. We can conclude that our data does not points to any
correlation between the values of $r_0$ and $\Gamma$.

\section{Discussion}
\label{sec:discussion}

\subsection{Possible selection bias ?}
\label{sec:4.1}

The derived values of $\Gamma$ span about one order of magnitude,
between $10^2 - 10^3$. This range is similar to the typical values of
$\Gamma$ inferred from other methods, as discussed above. The values of
$r_0$ span a wider range of roughly two orders of magnitude, between
$10^{6.5} - 10^{9.5}$~cm. These values are higher than the typically
assumed value of $r_0 \lesssim 10^7$~cm.

\begin{figure}
\plotone{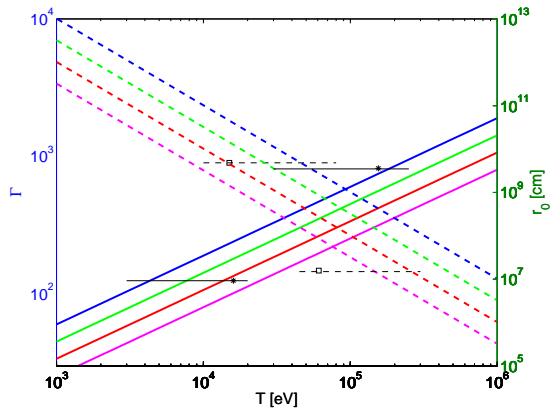}
\caption{Derived values of the Lorentz factor, $\Gamma$ (solid), and
  $r_0$ (dashed lines), as a function of the observed temperature and
  flux within typical detector's capabilities. The lines, from top to
  bottom, represent different GRB fluxes: $10^{-4}, 10^{-5}, 10^{-6},
  10^{-7}~{\rm erg~cm^{-2}~s^{-1}}$ (blue, greed, red, magenta,
  respectively). Lines derived for $F_{BB}/F_{tot} = 1/2$ and
  $z=1$. The stars show the location of the extremum observed values of
  $\Gamma$, and the squares the extremum observed values of $r_0$.  }
\label{fig:3}
\end{figure}

The values of $\Gamma$ and $r_0$ are derived from the identified
values of the observed temperature, $T^{ob.}$ as well as the fluxes
$F_{BB}^{ob.}$ and $F_{tot}^{ob.}$. Typical values of the observed GRB
fluxes are in the range of $F_{tot}^{ob.} \sim 10^{-7} - 10^{-4}~{\rm
  erg~cm^{-2}~s^{-1}}$ \citep[][see also Tables \ref{tab:T1},
  \ref{tab:T2}]{Kaneko+06, Gruber+14, VonKienlin+14}. The ratio of thermal to
total flux cannot be less than a few~\% (typically, it is of the order
of 50~\%, see Table \ref{tab:T2}), otherwise the thermal component
could not have been clearly identified. Furthermore, the BATSE
detector is sensitive at the range 20~keV - 2~MeV, while the GBM
detector has even broader sensitivity range, 8~keV - 40~MeV. Thus, a
thermal peak could be identified for the BATSE bursts at temperatures
$10^3 \lesssim T^{ob} \lesssim 10^6$~eV. Using these observational
constraints, we plot in Figure \ref{fig:3} the range of $r_0$ and
$\Gamma$ that could have been derived from observations of temperature
and fluxes in this range using Equations \ref{eq:1}, \ref{eq:2}, under
the assumption of $z=1$.

Clearly, the possible range of values of both $\Gamma$ and $r_0$ are
much greater than the observed ones. Values of $\Gamma$ are possible
in the range $10^{1.5} \lesssim \Gamma \lesssim 10^{3.5}$, and those
of $r_0$ could span a much broader range, $10^{5.5} \lesssim r_0
\lesssim 10^{13}$~cm (see Figure \ref{fig:3}). The addition of
uncertainty in the redshift would imply that, in principle, an even
broader range of parameter values could be obtained. However, this is
not observed: the inferred values of $\Gamma$ and $r_0$ span only a
small part of the range allowed by the detector's
capabilities. Combined with the fact that the histograms of both
$\Gamma$ and $r_0$ are close to Gaussian, we thus conclude that 
selection effects are likely to be ruled out.

Our sample is not homogeneous, as we use data obtained by two
different instruments (Fermi and BATSE detector). Furthermore, for the
GRBs in our category (I) and (II) we rely on analysis carried by
various authors. Thus, there is a risk of bias in the results. This
risk does not exist for GRBs in our category (III), as all the data is
taken from the BATSE instruments, and all the analysis was carried by
us. Our category (III) GRBs constitute the largest fraction of GRBs in
our sample ($36/47 \sim 75\%$).  When conducting a separate analysis
to the data in our category (III) GRBs only, we obtain similar results
to those obtained when analyzing the data in categories (I) and (II)
(see Figures \ref{fig:1} and \ref{fig:2}), indicating that selection
bias, if any, do not significantly affect the results.

An additional source of uncertainty is the unknown value of $Y \geq
1$, the ratio between energy released in the explosion and the energy
observed in $\gamma$-rays. As discussed above, the value we chose
here, $Y=2$, is a realistic approximation, based on (1) estimates from
the literature; (2) the fact that the values of $\Gamma$ obtained
using this value are consistent with measurements based on other
methods; and (3) the fact that a higher value would imply, in some
cases, values of $r_0$ lower than the gravitational radius, which is
unphysical. Nonetheless, we stress that as the derived values of both
$\Gamma$ and $r_0$ depend on $Y$ they are obviously sensitive to the
uncertainty in the value of this parameter.

\subsection{How ubiquitous is thermal emission in GRBs ?}
\label{sec:4.2}

The size of the sample considered in this work, 47 GRBs, is a small
fraction of the total number of GRBs observed to date. All GRBs in our
sample are characterized by a clearly detected, significant thermal
(Planck-like) component, which shows a well defined, repetitive
temporal behavior. We point out that a necessary condition for
performing this analysis is the ability to carry out a time-resolved
spectral analysis. This limits the number of GRBs in which this
analysis could be executed to only those that show relatively long,
smooth pulses, and, in addition, are bright. Furthermore, we point out
that the traditional use of a ``Band'' function on its own in fitting
the data, excludes the possibility of identifying a thermal component
on top of a non-thermal one, as the ``Band'' function is simply not
capable of capturing a thermal peak, where it exists. Thus, in order
to obtain a reliable estimate of the fraction of GRBs with a thermal
component, a re-analysis of the entire sample is needed. Recent works
\citep{Burgess+14, Guiriec+15} in which re-analysis of the data was
done, indicate that tens of \% of bright GRBs may show evidence for
the existence of a distinct thermal component. As shown in Table
\ref{tab:T1}, GRBs which do show evidence for a thermal component span
a wide range of properties in terms of fluxes, redshifts and light
curves, further supporting the idea that they are ubiquitous. 

The existence of a thermal component can be further deduced
indirectly, in those GRBs in which a clear ``Planck'' spectrum is not
observed. This is done by analyzing the low energy spectral slopes
obtained using the ``Band'' fits. As was shown already in 1998
\citep{Preece+98, Preece+02, GCG03}, the low energy spectral slopes of
over 85\% of the GRBs are too hard to be consistent with the
(optically thin) synchrotron emission. A more recent analysis
\citep{AB15, Yu+15b} reveals that the spectral width of nearly 100\%
of GRB pulses are inconsistent with having a synchrotron origin,
unless additional assumptions are made, such as variation of the
magnetic field with radius \citep{PZ06, UZ14, Zhang_BB+15}.

One possible solution that can explain the observed spectra of those
GRBs that do not show clear evidence of a distinct thermal component
(which are the vast majority) is that they do in fact originate from a
thermal component, that is modified due to sub-photospheric energy
dissipation \citep{PMR05, PMR06, Giannios08, Giannios12, VLP13,
  Lazzati+13, DZ14, KL14, GZ15, CL15, VB15, Ahlgren+15}. If this
hypothesis is correct, then the origin of the ``Band'' peak is
Comptonization of the thermal component. A similar mechanism may be
responsible for the non-thermal part of the spectra in those GRBs in
which a thermal component is pronounced, though an alternative
synchrotron model for the non-thermal part is also a possibility in
these bursts \citep{Burgess+11, Burgess+14, Yu+15}. Thus, in this
model, the key difference between GRBs in our sample (in which a
thermal component is identified) and other GRBs, is that for GRBs in
our sample the ``Planck'' part of the spectrum suffers only relatively
weak distortion. This could be due to the lack of significant
dissipation at small (sub-photospheric) radii. We point out that in
the classical fireball model, where internal shocks may be the main
source of energy dissipation, there is only a low limit constraint on
the location of the internal shocks, $r \gtrsim \Gamma^2 c \delta t$,
where $\delta t$ is the variability time. Thus, for smooth bursts, for
which $\delta t$ is relatively long, internal shocks occur far above
the photosphere \citep{RM05}. As a result, if indeed internal shocks
are the main cause of kinetic energy dissipation, one naturally
expects a pure photospheric component to be more pronounced in smooth
bursts. We stress, though, that so far no systematic study of
  possible correlation between the smoothness of lightcurves and the
  existence and properties of a thermal component was carried.

An alternative possibility of explaining the various spectra is due to
difference in magnetization \citep{ZP09, ZY11, BP14, GZ15, BP15}.  In
highly magnetized outflows (characterized by $\sigma = E_M / E_k \gg
1$, where $E_M$ is the energy stored in the magnetic field and $E_k$
is the kinetic energy), photospheric emission is both suppressed
\citep[e.g.,][]{ZP09} and is further shifted to higher energies due to
suppression of photon production below the photosphere. Thus,
according to this model, the difference between the GRBs in our sample
and other GRBs is that the outflow of the GRBs in our sample happen to
have a weaker magnetization, $\sigma \lesssim 1$. However, Currently,
there is no clear way to determine the magnetization of GRB outflows
without additional assumptions. On the other hand, we point out that
an additional assumptions, e.g., about the value of $r_0$ can be used
to interpret the data in the framework of the magnetized outflow
model, and use this interpretation to infer the value of the
magnetization parameter \citep[see][for details]{GZ15}.

Finally, we point out that the larger part of our sample (our category
III) are GRBs detected by BATSE. This is mainly due to the fact that a
systematic analysis in search for a thermal component was carried out,
so far, only on BATSE data, but not on data obtained by Fermi. It is
therefore not clear yet what fraction of GRBs detected by Fermi show
clear evidence for the existence of a thermal component.  A recent
analysis \citep{Guiriec+15b} indicates that the fraction of thermal
component in Fermi bursts may be similar to those observed in BATSE
bursts, though further analysis is still needed.

\subsection{Possible evidence of a massive progenitor}
\label{sec:4.3}

Within the framework of the ``collapsar'' model \citep{Woosley93}, the
jet drills its way through the collapsing star. For a freely expanding
outflow, it is typically assumed that the initial acceleration radius,
$r_0$ is close to the gravitational radius of the newly-formed black
hole, namely $r_0 \lesssim 10^7$~cm. However, the results presented
here indicate a rather different value of $\langle r_0 \rangle =
10^{8}$~cm, namely, about a factor 10 higher.

We should stress that the idea that $r_0$ should be $\approx$~few
gravitational radii is of theoretical origin, and is correct in the
classical (low magnetization) ``fireball'' model, in the limit of free
outflow. A small acceleration radius is associated, via light
crossing time argument, with short variability timescales, as $\delta t =
r_0/c$. However, the minimum observed variability time in GRBs is
$\gtrsim 10$~ms, with an average value of $\approx 500$~ms
\citep{GB14}. Variability timescales of 10~ms correspond to $r_0 \sim
10^{8.5}$~cm. Thus, in fact, there is no observational evidence
supporting values of $r_0$ much smaller than this value.

Adopting the basic framework of the ``collapsar'' model, 
the collimated outflow within the collapsing star cannot be
described as a free outflow. As the jet propagates through the star,
it forms a hot ``cocoon'' made of hot stellar material heated by the
jet itself that surrounds the jet. Furthermore, as the jet drills a
funnel through the stellar material, it is confined by the funnel
walls. The jet thermal pressure decreases with distance from the
origin. Thus, if the external pressure decreases slower than the jet
pressure, a recollimation shock must form. The shape of this
recollimation shock depends on the balance between the external
pressure and the jet ram pressure, and can be calculated analytically
\citep{KF97, NS09}.

This recollimation shock is very efficient in dissipating the kinetic
energy \citep{NS09, Nal12}. Thus, while the Lorentz factor increases
up to the recollimation shock radius, there is a sharp drop in the
outflow velocity as it encounters the recollimation shock to $\Gamma
\gtrsim 1$, before the outflow re-accelerates above this radius. This
is clearly seen in various numerical models of jet propagation inside
stellar cores \citep[e.g.,][]{Aloy+02, Morsony+07, MA09}, as well as
in analytical modeling \citep{Bromberg+11a}. The radii of the
recollimation shocks are typically seen at $\sim few \times 10^9$~cm
\citep{MA09, MI13, Lopez+13}, though in some of the simulations,
recollimation shocks can be identified only above $10^9$~cm due to
lack of numerical resolution at smaller radii.

Once the jet completes its crossing of the stellar core, both the jet
and the cocoon expand into the stellar envelope and into the
interstellar medium (ISM). At this stage, the external pressure
rapidly drops, and the radius of the recollimation shock gradually
increases, until the shocks eventually disappear and the flow becomes
free. However, this stage lasts typically for a duration of
$\gtrsim$~ten seconds (at least several sound crossing times);
\citep[e.g.,][]{Morsony+07, MA09, Lopez+13}. During this epoch, the
recollimation shock is roughly at its initial location. It can be
shown that the presence of a recollimation shock does change the
length scale of the jet ($r_0$) in certain conditions, especially if
the jet is brought back to causal contact by the collimation. In that
case, the measured value of $r_0$ would not correspond to the size of
the acceleration region nor to the radius at which the shocks happen,
but to $r_0 \Gamma_{\rm{sh}}$, where $\Gamma_{\rm{sh}}$ is the Lorentz
factor of the shocked jet. In our case, a Lorenz factor of $\sim30$
would be able to reconcile the observations with the theoretical
prediction of a small acceleration region of $\sim10^7$~cm. While
simulations show that typically $\Gamma_{\rm{sh}}<30$ \citep[see,
  e.g.,][]{Morsony+07, Lopez+13}, they do also show that shocks cause
mass entrainment on the jet and that the jet propagation is
non-conical at least for a few tens of seconds
\citep{Morsony+07}. Both these effect contribute also to causing a
measured value of $r_0$ that is larger than the size of the
acceleration region (Lazzati, Pe'er, Ryde, et al in preparation).

We may thus conclude that the results obtained here, of $r_0 \sim
  10^{7} - 10^{9}$~cm (despite all the uncertainties discussed above)
are consistent with the picture that emerges from the numerical
simulations, and that they provide indirect evidence that even the
brightest long duration bursts at cosmological redshifts can be produced
by the core-collapse of massive stars. Furthermore, this implies that
if indeed the progenitors of short GRBs are merger of binaries rather
than collapses of massive stars, then the values of $r_0$ are expected
to be much smaller in these bursts. While one of the bursts with a
high value of $r_0$ in our sample (GRB120323) is categorized as short,
we note that the discrimination between ``long'' and ``short'' is
statistical in nature, and thus it is possible that this burst may
originate from a collapse of a massive star, similar to classical long
GRBs \citep[see][]{Zhang+09}.


\section{Summary and conclusions}
\label{sec:summary}

In this work, we analyzed a sample of 47 GRBs which show significant
thermal emission component. We split our sample into three categories:
(I) 7 GRBs with known redshifts; (II) 4 GRBs detected by Fermi, in
which the redshifts are unknown; and (III) 36 GRBs detected by BATSE,
selected from the sample of \citet{RP09} based on clear identification
of the break time. We analyzed the dynamical properties of these GRBs
based on standard ``fireball'' model assumptions using the method
derived in \citet{Peer+07}. We found a mean Lorentz factor of $\langle
\Gamma \rangle = 370$, consistent with values obtained by different
methods. However, we further found that the acceleration radius $r_0$
ranges between $4 \times 10^6 \leq r_0 \leq 2 \times 10^9$~cm, with
mean value $\langle r_0 \rangle \simeq 10^8$~cm. This is $\approx$one
orders of magnitude above the commonly assumed value of $\sim
10^7$~cm. We found only a very weak correlation between the values of
$\Gamma$ and $r_0$, that cannot be used to put further constraints on
the properties of the progenitor. We showed that the derived values of
$r_0$ are only weakly sensitive to the uncertainty in the redshifts.

We showed in \S\ref{sec:4.1} that selection biases are unlikely to
affect our obtained results. We further argued (\S\ref{sec:4.2}) that
despite the fact that GRBs with a pure thermal emission component are
relatively rare, the existence of such a component is likely to be
very ubiquitous, and possibly it exists in nearly all GRBs, albeit in
most of them it is distorted. We claimed that within the context of
the standard fireball model, it is more likely to detect a pure
thermal component in long, smooth GRBs, that are less variable, and as
a result internal shocks occur above the photosphere. Finally, we
argued in \S\ref{sec:4.3},  that the most likely interpretation of
  the values of $r_0$ are propagation effects in massive progenitor
  stars, such as collimation shocks, mass entrainment, or non-conical
  expansion. 


An alternative scenario that can explain the relatively large values
of $r_0$ is that the acceleration in fact begins at radii much greater
than the gravitational radius of the newly formed black hole. This
could be due to initially large jet opening angle, followed by a
strong poloidal recollimation at $\sim 10^2 - 10^3~r_g$ that
accelerates the jet. Some evidence for a similar scenario exists in a
different object, the active galaxy M87 \citep{Junor+99}. This
scenario requires very strong poloidal magnetic fields, whose
existence in a GRB environment is uncertain.

The calculation of $r_0$ and $\Gamma$ in Equations \ref{eq:1},
\ref{eq:2} were done under the assumption of the ``classical'' (weakly
magnetized) ``fireball'' model. The values of $\Gamma$ that we found
are consistent with the values found using other methods, that are
independent of some of the underlying ``fireball'' model assumptions,
such as afterglow measurements \citep[e.g.,][]{Racusin+11}.  These
values are more than an order of magnitude higher than the values of
$\Gamma$ inferred in active galactic nuclei (AGNs) and X-ray
binaries. Thus, while magnetic acceleration \citep[e.g., via
  the][process]{BZ77} is likely to occur in XRBs and AGNs, currently
there is no evidence that a similar mechanism is at work in GRBs as
well.

In addition to the GRBs considered in this work, thermal emission was
found recently in several GRBs in the X-ray band, extending well into
the afterglow phase. These include both low luminosity GRBs such as
GRB060218 \citep{Campana+06, Shcherbakov+13} or GRB100316D
\citep{Starling+11} but also many GRBs with typical luminosities
\citep[e.g.,][]{Page+11, Starling+12, SpS12, FW13, Bellm+14,
  Schulze+14, Piro+14}.  Common to all GRBs in this category are (1)
the fact that the thermal component is observed well into the
afterglow phase; and (2) the inferred values of the Lorentz factors
are at least an order of magnitude lower than that of GRBs in our
sample: $\Gamma \lesssim$ few tens, and in some cases much lower,
$\Gamma \gtrsim 1$. These results indicate that most likely the
physical origin of the thermal component in bursts in this sample is
different than bursts in our sample, the leading models being
supernovae shock breakout and emission from the emerging cocoon. Given
this different origin, we excluded these bursts from the analysis
carried out here.

Finally, we stress that as the calculations here are carried under
the standard assumptions of the classical ``fireball'' model, any
inconsistency that may be found in the future between the values of
$r_0$ derived by our method and those derived in alternative ways
(e.g., via variability time argument) would question the validity of
the entire ``fireball'' model.

\acknowledgments 
AP wishes to thank Miguel Aloy for useful discussions. We would like
to thank Mohammed Basha, Kim Page and Motoko Serino for providing us
with direct access to their data. We would further like to thank the
referee, Dr. Bing Zhang for many useful comments, that helped us
improve this manuscript. This research was partially supported by the
European Union Seventh Framework Programme (FP7/2007-2013) under grant
agreement ${\rm n}^\circ$ 618499.  This work was supported in part by
National Science Foundation grant No. PHYS-1066293 and the hospitality
of the Aspen Center for Physics.

\bibliographystyle{/Users/apeer/Documents/Bib/apj}

\begin{thebibliography}{98}
\expandafter\ifx\csname natexlab\endcsname\relax\def\natexlab#1{#1}\fi

\bibitem[{{Abramowicz} {et~al.}(1991){Abramowicz}, {Novikov}, \&
  {Paczynski}}]{ANP91}
{Abramowicz}, M.~A., {Novikov}, I.~D., \& {Paczynski}, B. 1991, \apj, 369, 175

\bibitem[{{Ahlgren} {et~al.}(2015){Ahlgren}, {Larsson}, {Nymark}, {Ryde}, \&
  {Pe'er}}]{Ahlgren+15}
{Ahlgren}, B., {Larsson}, J., {Nymark}, T., {Ryde}, F., \& {Pe'er}, A. 2015,
  \mnras, 454, L31

\bibitem[{{Aloy} {et~al.}(2002){Aloy}, {Ib{\'a}{\~n}ez}, {Miralles}, \&
  {Urpin}}]{Aloy+02}
{Aloy}, M.-A., {Ib{\'a}{\~n}ez}, J.-M., {Miralles}, J.-A., \& {Urpin}, V. 2002,
  \aap, 396, 693

\bibitem[{{Axelsson} {et~al.}(2012){Axelsson}, {Baldini}, {Barbiellini},
  {Baring}, {Bellazzini}, {Bregeon}, {Brigida}, {Bruel}, {Buehler},
  {Caliandro}, {Cameron}, {Caraveo}, {Cecchi}, {Chaves}, {Chekhtman}, {Chiang},
  {Claus}, {Conrad}, {Cutini}, {D'Ammando}, {de Palma}, {Dermer}, {Silva},
  {Drell}, {Favuzzi}, {Fegan}, {Ferrara}, {Focke}, {Fukazawa}, {Fusco},
  {Gargano}, {Gasparrini}, {Gehrels}, {Germani}, {Giglietto}, {Giroletti},
  {Godfrey}, {Guiriec}, {Hadasch}, {Hanabata}, {Hayashida}, {Hou}, {Iyyani},
  {Jackson}, {Kocevski}, {Kuss}, {Larsson}, {Larsson}, {Longo}, {Loparco},
  {Lundman}, {Mazziotta}, {McEnery}, {Mizuno}, {Monzani}, {Moretti},
  {Morselli}, {Murgia}, {Nuss}, {Nymark}, {Ohno}, {Omodei}, {Pesce-Rollins},
  {Piron}, {Pivato}, {Racusin}, {Rain{\`o}}, {Razzano}, {Razzaque}, {Reimer},
  {Roth}, {Ryde}, {Sanchez}, {Sgr{\`o}}, {Siskind}, {Spandre}, {Spinelli},
  {Stamatikos}, {Tibaldo}, {Tinivella}, {Usher}, {Vandenbroucke}, {Vasileiou},
  {Vianello}, {Vitale}, {Waite}, {Winer}, {Wood}, {Burgess}, {Bhat},
  {Bissaldi}, {Briggs}, {Connaughton}, {Fishman}, {Fitzpatrick}, {Foley},
  {Gruber}, {Kippen}, {Kouveliotou}, {Jenke}, {McBreen}, {McGlynn}, {Meegan},
  {Paciesas}, {Pelassa}, {Preece}, {Tierney}, {von Kienlin}, {Wilson-Hodge},
  {Xiong}, \& {Pe'er}}]{Axelsson+12}
{Axelsson}, M., {Baldini}, L., {Barbiellini}, G., et. al. 2012, \apjl, 757, L31

\bibitem[{{Axelsson} \& {Borgonovo}(2015)}]{AB15}
{Axelsson}, M. \& {Borgonovo}, L. 2015, \mnras, 447, 3150

\bibitem[{{Basha}(2013)}]{Basha13}
{Basha}, M.~A.~F. 2013, \apss, 343, 107

\bibitem[{{B{\'e}gu{\'e}} \& {Pe'er}(2015)}]{BP15}
{B{\'e}gu{\'e}}, D. \& {Pe'er}, A. 2015, \apj, 802, 134

\bibitem[{{Bellm} {et~al.}(2014){Bellm}, {Barri{\`e}re}, {Bhalerao}, {Boggs},
  {Cenko}, {Christensen}, {Craig}, {Forster}, {Fryer}, {Hailey}, {Harrison},
  {Horesh}, {Kouveliotou}, {Madsen}, {Miller}, {Ofek}, {Perley}, {Rana},
  {Reynolds}, {Stern}, {Tomsick}, \& {Zhang}}]{Bellm+14}
{Bellm}, E.~C., {Barri{\`e}re}, N.~M., {Bhalerao}, V., et. al., 
  2014, \apjl, 784, L19

\bibitem[{{Beniamini} \& {Piran}(2014)}]{BP14}
{Beniamini}, P. \& {Piran}, T. 2014, \mnras, 445, 3892

\bibitem[{{Blandford} \& {Znajek}(1977)}]{BZ77}
{Blandford}, R.~D. \& {Znajek}, R.~L. 1977, \mnras, 179, 433

\bibitem[{{Bromberg} {et~al.}(2011){Bromberg}, {Nakar}, {Piran}, \&
  {Sari}}]{Bromberg+11a}
{Bromberg}, O., {Nakar}, E., {Piran}, T., \& {Sari}, R. 2011, \apj, 740, 100

\bibitem[{{Burgess} {et~al.}(2011){Burgess}, {Preece}, {Baring}, {Briggs},
  {Connaughton}, {Guiriec}, {Paciesas}, {Meegan}, {Bhat}, {Bissaldi}, \& et.
  al.}]{Burgess+11}
{Burgess}, J.~M., {Preece}, R.~D., {Baring}, M.~G., et. al. 2011, \apj, 741, 24

\bibitem[{{Burgess} {et~al.}(2014){Burgess}, {Preece}, {Connaughton}, {Briggs},
  {Goldstein}, {Bhat}, {Greiner}, {Gruber}, {Kienlin}, {Kouveliotou},
  {McGlynn}, {Meegan}, {Paciesas}, {Rau}, {Xiong}, {Axelsson}, {Baring},
  {Dermer}, {Iyyani}, {Kocevski}, {Omodei}, {Ryde}, \& {Vianello}}]{Burgess+14}
{Burgess}, J.~M., {Preece}, R.~D., {Connaughton}, V., et. al. 2014, \apj,
  784, 17

\bibitem[{{Campana} {et~al.}(2006){Campana}, {Mangano}, {Blustin}, {Brown},
  {Burrows}, {Chincarini}, {Cummings}, {Cusumano}, {Della Valle}, {Malesani},
  {M{\'e}sz{\'a}ros}, {Nousek}, {Page}, {Sakamoto}, {Waxman}, {Zhang}, {Dai},
  {Gehrels}, {Immler}, {Marshall}, {Mason}, {Moretti}, {O'Brien}, {Osborne},
  {Page}, {Romano}, {Roming}, {Tagliaferri}, {Cominsky}, {Giommi}, {Godet},
  {Kennea}, {Krimm}, {Angelini}, {Barthelmy}, {Boyd}, {Palmer}, {Wells}, \&
  {White}}]{Campana+06}
{Campana}, S., {Mangano}, V., {Blustin}, A.~J., et. al. 2006, \nat, 442, 1008

\bibitem[{{Cenko} {et~al.}(2011){Cenko}, {Frail}, {Harrison}, {Haislip},
  {Reichart}, {Butler}, {Cobb}, {Cucchiara}, {Berger}, {Bloom}, {Chandra},
  {Fox}, {Perley}, {Prochaska}, {Filippenko}, {Glazebrook}, {Ivarsen},
  {Kasliwal}, {Kulkarni}, {LaCluyze}, {Lopez}, {Morgan}, {Pettini}, \&
  {Rana}}]{Cenko+11}
{Cenko}, S.~B., {Frail}, D.~A., {Harrison}, F.~A., et. al. 2011, \apj, 732, 29

\bibitem[{{Chhotray} \& {Lazzati}(2015)}]{CL15}
{Chhotray}, A. \& {Lazzati}, D. 2015, \apj, 802, 132

\bibitem[{{Deng} \& {Zhang}(2014)}]{DZ14}
{Deng}, W. \& {Zhang}, B. 2014, \apj, 785, 112

\bibitem[{{Friis} \& {Watson}(2013)}]{FW13}
{Friis}, M. \& {Watson}, D. 2013, \apj, 771, 15

\bibitem[{{Gao} \& {Zhang}(2015)}]{GZ15}
{Gao}, H. \& {Zhang}, B. 2015, \apj, 801, 103

\bibitem[{{Ghirlanda} {et~al.}(2003){Ghirlanda}, {Celotti}, \&
  {Ghisellini}}]{GCG03}
{Ghirlanda}, G., {Celotti}, A., \& {Ghisellini}, G. 2003, \aap, 406, 879

\bibitem[{{Ghirlanda} {et~al.}(2013){Ghirlanda}, {Pescalli}, \&
  {Ghisellini}}]{Ghirlanda+13}
{Ghirlanda}, G., {Pescalli}, A., \& {Ghisellini}, G. 2013, \mnras, 432, 3237

\bibitem[{{Giannios}(2008)}]{Giannios08}
{Giannios}, D. 2008, \aap, 480, 305

\bibitem[{{Giannios}(2012)}]{Giannios12}
---. 2012, \mnras, 422, 3092

\bibitem[{{Golkhou} \& {Butler}(2014)}]{GB14}
{Golkhou}, V.~Z. \& {Butler}, N.~R. 2014, \apj, 787, 90

\bibitem[{{Gruber} {et~al.}(2014){Gruber}, {Goldstein}, {Weller von Ahlefeld},
  {Narayana Bhat}, {Bissaldi}, {Briggs}, {Byrne}, {Cleveland}, {Connaughton},
  {Diehl}, {Fishman}, {Fitzpatrick}, {Foley}, {Gibby}, {Giles}, {Greiner},
  {Guiriec}, {van der Horst}, {von Kienlin}, {Kouveliotou}, {Layden}, {Lin},
  {Meegan}, {McGlynn}, {Paciesas}, {Pelassa}, {Preece}, {Rau}, {Wilson-Hodge},
  {Xiong}, {Younes}, \& {Yu}}]{Gruber+14}
{Gruber}, D., {Goldstein}, A., {Weller von Ahlefeld}, V., et. al. 2014, \apjs, 211, 12

\bibitem[{{Guiriec} {et~al.}(2011){Guiriec}, {Connaughton}, {Briggs},
  {Burgess}, {Ryde}, {Daigne}, {M{\'e}sz{\'a}ros}, {Goldstein}, {McEnery},
  {Omodei}, {Bhat}, {Bissaldi}, {Camero-Arranz}, {Chaplin}, {Diehl}, {Fishman},
  {Foley}, {Gibby}, {Giles}, {Greiner}, {Gruber}, {von Kienlin}, {Kippen},
  {Kouveliotou}, {McBreen}, {Meegan}, {Paciesas}, {Preece}, {Rau}, {Tierney},
  {van der Horst}, \& {Wilson-Hodge}}]{Guiriec+11}
{Guiriec}, S., {Connaughton}, V., {Briggs}, M.~S., et. al. 2011, \apjl,
  727, L33

\bibitem[{{Guiriec} {et~al.}(2013){Guiriec}, {Daigne}, {Hasco{\"e}t},
  {Vianello}, {Ryde}, {Mochkovitch}, {Kouveliotou}, {Xiong}, {Bhat}, {Foley},
  {Gruber}, {Burgess}, {McGlynn}, {McEnery}, \& {Gehrels}}]{Guiriec+13}
{Guiriec}, S., {Daigne}, F., {Hasco{\"e}t}, R., et. al. 2013, \apj, 770, 32

\bibitem[{{Guiriec} {et~al.}(2015{\natexlab{a}}){Guiriec}, {Gonzalez},
  {Sacahui}, {Kouveliotou}, {Gehrels}, \& {McEnery}}]{Guiriec+15b}
{Guiriec}, S., {Gonzalez}, M.~M., {Sacahui}, J.~R., {Kouveliotou}, C.,
  {Gehrels}, N., \& {McEnery}, J. 2015{\natexlab{a}}, ArXiv e-prints

\bibitem[{{Guiriec} {et~al.}(2015{\natexlab{b}}){Guiriec}, {Kouveliotou},
  {Daigne}, {Zhang}, {Hasco{\"e}t}, {Nemmen}, {Thompson}, {Bhat}, {Gehrels},
  {Gonzalez}, {Kaneko}, {McEnery}, {Mochkovitch}, {Racusin}, {Ryde}, {Sacahui},
  \& {{\"U}nsal}}]{Guiriec+15}
{Guiriec}, S., {Kouveliotou}, C., {Daigne}, F., et. al. 2015{\natexlab{b}}, \apj,
  807, 148

\bibitem[{{Hasco{\"e}t} {et~al.}(2013){Hasco{\"e}t}, {Daigne}, \&
  {Mochkovitch}}]{Hascoet+13}
{Hasco{\"e}t}, R., {Daigne}, F., \& {Mochkovitch}, R. 2013, \aap, 551, A124

\bibitem[{{Iyyani} {et~al.}(2015){Iyyani}, {Ryde}, {Ahlgren}, {Burgess},
  {Larsson}, {Pe'er}, {Lundman}, {Axelsson}, \& {McGlynn}}]{Iyyani+15}
{Iyyani}, S., {Ryde}, F., {Ahlgren}, B., et. al. 2015, \mnras,
  450, 1651

\bibitem[{{Iyyani} {et~al.}(2013){Iyyani}, {Ryde}, {Axelsson}, {Burgess},
  {Guiriec}, {Larsson}, {Lundman}, {Moretti}, {McGlynn}, {Nymark}, \&
  {Rosquist}}]{Iyyani+13}
{Iyyani}, S., {Ryde}, F., {Axelsson}, M., et. al. 2013, \mnras, 433, 2739

\bibitem[{{Jakobsson} {et~al.}(2006){Jakobsson}, {Levan}, {Fynbo}, {Priddey},
  {Hjorth}, {Tanvir}, {Watson}, {Jensen}, {Sollerman}, {Natarajan},
  {Gorosabel}, {Castro Cer{\'o}n}, {Pedersen}, {Pursimo}, {{\'A}rnad{\'o}ttir},
  {Castro-Tirado}, {Davis}, {Deeg}, {Fiuza}, {Mikolaitis}, \&
  {Sousa}}]{Jakobsson+06}
{Jakobsson}, P., {Levan}, A., {Fynbo}, J.~P.~U., et. al. 2006,
  \aap, 447, 897

\bibitem[{{Junor} {et~al.}(1999){Junor}, {Biretta}, \& {Livio}}]{Junor+99}
{Junor}, W., {Biretta}, J.~A., \& {Livio}, M. 1999, \nat, 401, 891

\bibitem[{{Kaneko} {et~al.}(2006){Kaneko}, {Preece}, {Briggs}, {Paciesas},
  {Meegan}, \& {Band}}]{Kaneko+06}
{Kaneko}, Y., {Preece}, R.~D., {Briggs}, M.~S., {Paciesas}, W.~S., {Meegan},
  C.~A., \& {Band}, D.~L. 2006, \apjs, 166, 298

\bibitem[{{Keren} \& {Levinson}(2014)}]{KL14}
{Keren}, S. \& {Levinson}, A. 2014, \apj, 789, 128

\bibitem[{{Komissarov} \& {Falle}(1997)}]{KF97}
{Komissarov}, S.~S. \& {Falle}, S.~A.~E.~G. 1997, \mnras, 288, 833

\bibitem[{{Krolik} \& {Pier}(1991)}]{KP91}
{Krolik}, J.~H. \& {Pier}, E.~A. 1991, \apj, 373, 277

\bibitem[{{Kumar} \& {Zhang}(2014)}]{KZ14}
{Kumar}, P. \& {Zhang}, B. 2014, \physrep, 561, 1

\bibitem[{{Larsson} {et~al.}(2015){Larsson}, {Racusin}, \&
  {Burgess}}]{Larsson+15}
{Larsson}, J., {Racusin}, J.~L., \& {Burgess}, J.~M. 2015, \apjl, 800, L34

\bibitem[{{Lazzati} {et~al.}(2013){Lazzati}, {Morsony}, {Margutti}, \&
  {Begelman}}]{Lazzati+13}
{Lazzati}, D., {Morsony}, B.~J., {Margutti}, R., \& {Begelman}, M.~C. 2013,
  \apj, 765, 103

\bibitem[{{Liang} {et~al.}(2010){Liang}, {Yi}, {Zhang}, {L{\"u}}, {Zhang}, \&
  {Zhang}}]{Liang+10}
{Liang}, E.-W., {Yi}, S.-X., {Zhang}, J., {L{\"u}}, H.-J., {Zhang}, B.-B., \&
  {Zhang}, B. 2010, \apj, 725, 2209

\bibitem[{{Lithwick} \& {Sari}(2001)}]{LS01}
{Lithwick}, Y. \& {Sari}, R. 2001, \apj, 555, 540

\bibitem[{{L{\'o}pez-C{\'a}mara} {et~al.}(2013){L{\'o}pez-C{\'a}mara},
  {Morsony}, {Begelman}, \& {Lazzati}}]{Lopez+13}
{L{\'o}pez-C{\'a}mara}, D., {Morsony}, B.~J., {Begelman}, M.~C., \& {Lazzati},
  D. 2013, \apj, 767, 19

\bibitem[{{Lundman} {et~al.}(2013){Lundman}, {Pe'er}, \& {Ryde}}]{LPR13}
{Lundman}, C., {Pe'er}, A., \& {Ryde}, F. 2013, \mnras, 428, 2430

\bibitem[{{Margutti} {et~al.}(2013){Margutti}, {Soderberg}, {Wieringa},
  {Edwards}, {Chevalier}, {Morsony}, {Barniol Duran}, {Sironi}, {Zauderer},
  {Milisavljevic}, {Kamble}, \& {Pian}}]{Margutti+13}
{Margutti}, R., {Soderberg}, A.~M., {Wieringa}, M.~H., et. al. 2013,
  \apj, 778, 18

\bibitem[{{McGlynn} \& {Fermi GBM Collaboration}(2012)}]{McGlynn+12}
{McGlynn}, S. \& {Fermi GBM Collaboration}. 2012, in $\gamma$-Ray Bursts 2012
  Conference (GRB 2012), 12

\bibitem[{{M{\'e}sz{\'a}ros}(2006)}]{Meszaros06}
{M{\'e}sz{\'a}ros}, P. 2006, Reports on Progress in Physics, 69, 2259

\bibitem[{{M{\'e}sz{\'a}ros} \& {Rees}(2000)}]{MR00}
{M{\'e}sz{\'a}ros}, P. \& {Rees}, M.~J. 2000, \apj, 530, 292

\bibitem[{{Mizuta} \& {Aloy}(2009)}]{MA09}
{Mizuta}, A. \& {Aloy}, M.~A. 2009, \apj, 699, 1261

\bibitem[{{Mizuta} \& {Ioka}(2013)}]{MI13}
{Mizuta}, A. \& {Ioka}, K. 2013, \apj, 777, 162

\bibitem[{{Morsony} {et~al.}(2007){Morsony}, {Lazzati}, \&
  {Begelman}}]{Morsony+07}
{Morsony}, B.~J., {Lazzati}, D., \& {Begelman}, M.~C. 2007, \apj, 665, 569

\bibitem[{{Nalewajko}(2012)}]{Nal12}
{Nalewajko}, K. 2012, \mnras, 420, L48

\bibitem[{{Nalewajko} \& {Sikora}(2009)}]{NS09}
{Nalewajko}, K. \& {Sikora}, M. 2009, \mnras, 392, 1205

\bibitem[{{Paczynski}(1986)}]{Pac86}
{Paczynski}, B. 1986, \apjl, 308, L43

\bibitem[{{Paczynski}(1990)}]{Pac90}
---. 1990, \apj, 363, 218

\bibitem[{{Page} {et~al.}(2011){Page}, {Starling}, {Fitzpatrick}, {Pandey},
  {Osborne}, {Schady}, {McBreen}, {Campana}, {Ukwatta}, {Pagani}, {Beardmore},
  \& {Evans}}]{Page+11}
{Page}, K.~L., {Starling}, R.~L.~C., {Fitzpatrick}, G., et. al. 2011, \mnras, 416,
  2078

\bibitem[{{Page} {et~al.}(2009){Page}, {Willingale}, {Bissaldi}, {Postigo},
  {Holland}, {McBreen}, {O'Brien}, {Osborne}, {Prochaska}, {Rol}, {Rykoff},
  {Starling}, {Tanvir}, {van der Horst}, {Wiersema}, {Zhang}, {Aceituno},
  {Akerlof}, {Beardmore}, {Briggs}, {Burrows}, {Castro-Tirado}, {Connaughton},
  {Evans}, {Fynbo}, {Gehrels}, {Guidorzi}, {Howard}, {Kennea}, {Kouveliotou},
  {Pagani}, {Preece}, {Perley}, {Steele}, \& {Yuan}}]{Page+09}
{Page}, K.~L., {Willingale}, R., {Bissaldi}, E., et. al. 2009, \mnras, 400, 134

\bibitem[{{Pe'er}(2008)}]{Peer08}
{Pe'er}, A. 2008, \apj, 682, 463

\bibitem[{{Pe'er} {et~al.}(2005){Pe'er}, {M{\'e}sz{\'a}ros}, \& {Rees}}]{PMR05}
{Pe'er}, A., {M{\'e}sz{\'a}ros}, P., \& {Rees}, M.~J. 2005, \apj, 635, 476

\bibitem[{{Pe'er} {et~al.}(2006){Pe'er}, {M{\'e}sz{\'a}ros}, \& {Rees}}]{PMR06}
---. 2006, \apj, 642, 995

\bibitem[{{Pe'er} \& {Ryde}(2011)}]{PR11}
{Pe'er}, A. \& {Ryde}, F. 2011, \apj, 732, 49

\bibitem[{{Pe'er} {et~al.}(2007){Pe'er}, {Ryde}, {Wijers}, {M{\'e}sz{\'a}ros},
  \& {Rees}}]{Peer+07}
{Pe'er}, A., {Ryde}, F., {Wijers}, R.~A.~M.~J., {M{\'e}sz{\'a}ros}, P., \&
  {Rees}, M.~J. 2007, \apjl, 664, L1

\bibitem[{{Pe'er} \& {Zhang}(2006)}]{PZ06}
{Pe'er}, A. \& {Zhang}, B. 2006, \apj, 653, 454

\bibitem[{{Pe'er} {et~al.}(2012){Pe'er}, {Zhang}, {Ryde}, {McGlynn}, {Zhang},
  {Preece}, \& {Kouveliotou}}]{Peer+12}
{Pe'er}, A., {Zhang}, B.-B., {Ryde}, F., {McGlynn}, S., {Zhang}, B., {Preece},
  R.~D., \& {Kouveliotou}, C. 2012, \mnras, 420, 468

\bibitem[{{Piro} {et~al.}(2014){Piro}, {Troja}, {Gendre}, {Ghisellini},
  {Ricci}, {Bannister}, {Fiore}, {Kidd}, {Piranomonte}, \&
  {Wieringa}}]{Piro+14}
{Piro}, L., {Troja}, E., {Gendre}, B., et. al. 2014, \apjl, 790, L15

\bibitem[{{Preece} {et~al.}(2002){Preece}, {Briggs}, {Giblin}, {Mallozzi},
  {Pendleton}, {Paciesas}, \& {Band}}]{Preece+02}
{Preece}, R.~D., {Briggs}, M.~S., {Giblin}, T.~W., {Mallozzi}, R.~S.,
  {Pendleton}, G.~N., {Paciesas}, W.~S., \& {Band}, D.~L. 2002, \apj, 581, 1248

\bibitem[{{Preece} {et~al.}(1998){Preece}, {Briggs}, {Mallozzi}, {Pendleton},
  {Paciesas}, \& {Band}}]{Preece+98}
{Preece}, R.~D., {Briggs}, M.~S., {Mallozzi}, R.~S., {Pendleton}, G.~N.,
  {Paciesas}, W.~S., \& {Band}, D.~L. 1998, \apjl, 506, L23

\bibitem[{{Racusin} {et~al.}(2011){Racusin}, {Oates}, {Schady}, {Burrows}, {de
  Pasquale}, {Donato}, {Gehrels}, {Koch}, {McEnery}, {Piran}, {Roming},
  {Sakamoto}, {Swenson}, {Troja}, {Vasileiou}, {Virgili}, {Wanderman}, \&
  {Zhang}}]{Racusin+11}
{Racusin}, J.~L., {Oates}, S.~R., {Schady}, P., et. al. 2011, \apj,
  738, 138

\bibitem[{{Rees} \& {Meszaros}(1992)}]{RM92}
{Rees}, M.~J. \& {Meszaros}, P. 1992, \mnras, 258, 41P

\bibitem[{{Rees} \& {Meszaros}(1994)}]{RM94}
---. 1994, \apjl, 430, L93

\bibitem[{{Rees} \& {M{\'e}sz{\'a}ros}(2005)}]{RM05}
{Rees}, M.~J. \& {M{\'e}sz{\'a}ros}, P. 2005, \apj, 628, 847

\bibitem[{{Ryde}(2004)}]{Ryde04}
{Ryde}, F. 2004, \apj, 614, 827

\bibitem[{{Ryde}(2005)}]{Ryde05}
---. 2005, \apjl, 625, L95

\bibitem[{{Ryde} {et~al.}(2010){Ryde}, {Axelsson}, {Zhang}, {McGlynn}, {Pe'er},
  {Lundman}, {Larsson}, {Battelino}, {Zhang}, {Bissaldi}, {Bregeon}, {Briggs},
  {Chiang}, {de Palma}, {Guiriec}, {Larsson}, {Longo}, {McBreen}, {Omodei},
  {Petrosian}, {Preece}, \& {van der Horst}}]{Ryde+10}
{Ryde}, F., {Axelsson}, M., {Zhang}, B.~B., et. al. 2010, \apjl, 709, L172

\bibitem[{{Ryde} \& {Pe'er}(2009)}]{RP09}
{Ryde}, F. \& {Pe'er}, A. 2009, \apj, 702, 1211

\bibitem[{{Santana} {et~al.}(2014){Santana}, {Barniol Duran}, \&
  {Kumar}}]{SBK14}
{Santana}, R., {Barniol Duran}, R., \& {Kumar}, P. 2014, \apj, 785, 29

\bibitem[{{Sari} \& {Piran}(1999)}]{SP99}
{Sari}, R. \& {Piran}, T. 1999, \apjl, 517, L109

\bibitem[{{Schulze} {et~al.}(2014){Schulze}, {Malesani}, {Cucchiara}, {Tanvir},
  {Kr{\"u}hler}, {de Ugarte Postigo}, {Leloudas}, {Lyman}, {Bersier},
  {Wiersema}, {Perley}, {Schady}, {Gorosabel}, {Anderson}, {Castro-Tirado},
  {Cenko}, {De Cia}, {Ellerbroek}, {Fynbo}, {Greiner}, {Hjorth}, {Kann},
  {Kaper}, {Klose}, {Levan}, {Mart{\'{\i}}n}, {O'Brien}, {Page}, {Pignata},
  {Rapaport}, {S{\'a}nchez-Ram{\'{\i}}rez}, {Sollerman}, {Smith}, {Sparre},
  {Th{\"o}ne}, {Watson}, {Xu}, {Bauer}, {Bayliss}, {Bj{\"o}rnsson}, {Bremer},
  {Cano}, {Covino}, {D'Elia}, {Frail}, {Geier}, {Goldoni}, {Hartoog},
  {Jakobsson}, {Korhonen}, {Lee}, {Milvang-Jensen}, {Nardini}, {Nicuesa
  Guelbenzu}, {Oguri}, {Pandey}, {Petitpas}, {Rossi}, {Sandberg}, {Schmidl},
  {Tagliaferri}, {Tilanus}, {Winters}, {Wright}, \& {Wuyts}}]{Schulze+14}
{Schulze}, S., {Malesani}, D., {Cucchiara}, A., et. al. 2014,
  \aap, 566, A102

\bibitem[{{Serino} {et~al.}(2011){Serino}, {Yoshida}, {Kawai}, {Nakagawa},
  {Mihara}, {Ueda}, {Nakahira}, {Eguchi}, {Hiroi}, {Ishikawa}, {Isobe},
  {Kimura}, {Kitayama}, {Kohama}, {Matsumura}, {Matsuoka}, {Morii}, {Nakajima},
  {Negoro}, {Shidatsu}, {Sootome}, {Sugimori}, {Sugizaki}, {Suwa}, {Toizumi},
  {Tomida}, {Tsuboi}, {Tsunemi}, {Ueno}, {Usui}, {Yamamoto}, {Yamaoka},
  {Yamauchi}, \& {Yamazaki}}]{Serino+11}
{Serino}, M., {Yoshida}, A., {Kawai}, N., et. al. 2011, \pasj,
  63, 1035

\bibitem[{{Shcherbakov} {et~al.}(2013){Shcherbakov}, {Pe'er}, {Reynolds},
  {Haas}, {Bode}, \& {Laguna}}]{Shcherbakov+13}
{Shcherbakov}, R.~V., {Pe'er}, A., {Reynolds}, C.~S., {Haas}, R., {Bode}, T.,
  \& {Laguna}, P. 2013, \apj, 769, 85

\bibitem[{{Sparre} \& {Starling}(2012)}]{SpS12}
{Sparre}, M. \& {Starling}, R.~L.~C. 2012, \mnras, 427, 2965

\bibitem[{{Starling} {et~al.}(2012){Starling}, {Page}, {Pe'Er}, {Beardmore}, \&
  {Osborne}}]{Starling+12}
{Starling}, R.~L.~C., {Page}, K.~L., {Pe'er}, A., {Beardmore}, A.~P., \&
  {Osborne}, J.~P. 2012, \mnras, 427, 2950

\bibitem[{{Starling} {et~al.}(2011){Starling}, {Wiersema}, {Levan}, {Sakamoto},
  {Bersier}, {Goldoni}, {Oates}, {Rowlinson}, {Campana}, {Sollerman}, {Tanvir},
  {Malesani}, {Fynbo}, {Covino}, {D'Avanzo}, {O'Brien}, {Page}, {Osborne},
  {Vergani}, {Barthelmy}, {Burrows}, {Cano}, {Curran}, {de Pasquale}, {D'Elia},
  {Evans}, {Flores}, {Fruchter}, {Garnavich}, {Gehrels}, {Gorosabel}, {Hjorth},
  {Holland}, {van der Horst}, {Hurkett}, {Jakobsson}, {Kamble}, {Kouveliotou},
  {Kuin}, {Kaper}, {Mazzali}, {Nugent}, {Pian}, {Stamatikos}, {Th{\"o}ne}, \&
  {Woosley}}]{Starling+11}
{Starling}, R.~L.~C., {Wiersema}, K., {Levan}, A.~J., et. al. 2011, \mnras, 411, 2792

\bibitem[{{Uhm} \& {Zhang}(2014)}]{UZ14}
{Uhm}, Z.~L. \& {Zhang}, B. 2014, Nature Physics, 10, 351

\bibitem[{{von Kienlin} {et~al.}(2014){von Kienlin}, {Meegan}, {Paciesas},
  {Bhat}, {Bissaldi}, {Briggs}, {Burgess}, {Byrne}, {Chaplin}, {Cleveland},
  {Connaughton}, {Collazzi}, {Fitzpatrick}, {Foley}, {Gibby}, {Giles},
  {Goldstein}, {Greiner}, {Gruber}, {Guiriec}, {van der Horst}, {Kouveliotou},
  {Layden}, {McBreen}, {McGlynn}, {Pelassa}, {Preece}, {Rau}, {Tierney},
  {Wilson-Hodge}, {Xiong}, {Younes}, \& {Yu}}]{VonKienlin+14}
{von Kienlin}, A., {Meegan}, C.~A., {Paciesas}, W.~S., et. al. 2014, \apjs, 211, 13

\bibitem[{{Vurm} \& {Beloborodov}(2015)}]{VB15}
{Vurm}, I. \& {Beloborodov}, A.~M. 2015, ArXiv e-prints

\bibitem[{{Vurm} {et~al.}(2013){Vurm}, {Lyubarsky}, \& {Piran}}]{VLP13}
{Vurm}, I., {Lyubarsky}, Y., \& {Piran}, T. 2013, \apj, 764, 143

\bibitem[{{Wang} {et~al.}(2015){Wang}, {Zhang}, {Liang}, {Gao}, {Li}, {Deng},
  {Qin}, {Tang}, {Kann}, {Ryde}, \& {Kumar}}]{Wang+15}
{Wang}, X.-G., {Zhang}, B., {Liang}, E.-W., et. al. 2015,
  \apjs, 219, 9

\bibitem[{{Woods} \& {Loeb}(1995)}]{WL95}
{Woods}, E. \& {Loeb}, A. 1995, \apj, 453, 583

\bibitem[{{Woosley}(1993)}]{Woosley93}
{Woosley}, S.~E. 1993, \apj, 405, 273

\bibitem[{{Wygoda} {et~al.}(2015){Wygoda}, {Guetta}, {Mandich}, \&
  {Waxman}}]{Wygoda+15}
{Wygoda}, N., {Guetta}, D., {Mandich}, M.-A., \& {Waxman}, E. 2015, ArXiv
  e-prints

\bibitem[{{Yu} {et~al.}(2015{\natexlab{a}}){Yu}, {Greiner}, {van Eerten},
  {Burgess}, {Narayana Bhat}, {Briggs}, {Connaughton}, {Diehl}, {Goldstein},
  {Gruber}, {Jenke}, {von Kienlin}, {Kouveliotou}, {Paciesas}, {Pelassa},
  {Preece}, {Roberts}, \& {Zhang}}]{Yu+15}
{Yu}, H.-F., {Greiner}, J., {van Eerten}, H., et. al. 2015{\natexlab{a}}, \aap, 573, A81

\bibitem[{{Yu} {et~al.}(2015{\natexlab{b}}){Yu}, {van Eerten}, {Greiner},
  {Sari}, {Narayana Bhat}, {von Kienlin}, {Paciesas}, \& {Preece}}]{Yu+15b}
{Yu}, H.-F., {van Eerten}, H.~J., {Greiner}, J., {Sari}, R., {Narayana Bhat},
  P., {von Kienlin}, A., {Paciesas}, W.~S., \& {Preece}, R.~D.
  2015{\natexlab{b}}, ArXiv e-prints

\bibitem[{{Zhang} \& {Pe'er}(2009)}]{ZP09}
{Zhang}, B. \& {Pe'er}, A. 2009, \apjl, 700, L65

\bibitem[{{Zhang} \& {Yan}(2011)}]{ZY11}
{Zhang}, B. \& {Yan}, H. 2011, \apj, 726, 90

\bibitem[{{Zhang} {et~al.}(2009){Zhang}, {Zhang}, {Virgili}, {Liang}, {Kann},
  {Wu}, {Proga}, {Lv}, {Toma}, {M{\'e}sz{\'a}ros}, {Burrows}, {Roming}, \&
  {Gehrels}}]{Zhang+09}
{Zhang}, B., {Zhang}, B.-B., {Virgili}, F.~J., et. al. 2009, \apj, 703, 1696

\bibitem[{{Zhang} {et~al.}(2015){Zhang}, {Uhm}, {Connaughton}, {Briggs}, \&
  {Zhang}}]{Zhang_BB+15}
{Zhang}, B.-B., {Uhm}, Z.~L., {Connaughton}, V., {Briggs}, M.~S., \& {Zhang},
  B. 2015, ArXiv e-prints

\end{thebibliography}


\end{document}